\documentclass[twocolumn]{aastex631}

\usepackage{amsmath}

\usepackage{caption}
\usepackage{subcaption}

\begin{document}

\title{The sensitivity of GPz estimates of photo-z posterior PDFs \\
to realistically complex training set imperfections}

\author{Natalia Stylianou}
\affiliation{School of Physics and Astronomy, University of Leicester, University Road, LE1 7RH, UK}

\author{Alex I. Malz}
\affiliation{German Centre for Cosmological Lensing, Astronomisches Institut, Ruhr-Universit{\"a}t Bochum, Universit{\"a}tsstr. 150, 44801 Bochum, Germany}

\author{Peter Hatfield}
\affiliation{Astrophysics, University of Oxford, Denys Wilkinson Building, Keble Road, Oxford OX1 3RH, UK}

\author{John Franklin Crenshaw}
\affiliation{DIRAC Institute and Department of Physics, University of Washington, Seattle, WA 98195, USA}

\author{Julia Gschwend}
\affiliation{Laboratório Interinstitucional de e-Astronomia (LIneA), Rua General José Cristino, 77, Rio de Janeiro, Brazil}

\begin{abstract}

The accurate estimation of photometric redshifts is crucial to many upcoming galaxy surveys, for example the Vera C. Rubin Observatory Legacy Survey of Space and Time (LSST). Almost all Rubin extragalactic and cosmological science requires accurate and precise calculation of photometric redshifts; many diverse approaches to this problem are currently in the process of being developed, validated, and tested. In this work, we use the photometric redshift code GPz to examine two realistically complex training set imperfections scenarios for machine learning based photometric redshift calculation: i) where the spectroscopic training set has a very different distribution in colour-magnitude space to the test set, and ii) where the effect of emission line confusion causes a fraction of the training spectroscopic sample to not have the true redshift. By evaluating the sensitivity of GPz to a range of increasingly severe imperfections, with a range of metrics (both of photo-z point estimates as well as posterior probability distribution functions, PDFs), we quantify the degree to which predictions get worse with higher degrees of degradation. In particular we find that there is a substantial drop-off in photo-z quality when line-confusion goes above $\sim1$\%, and sample incompleteness below a redshift of 1.5, for an experimental setup using data from the Buzzard Flock synthetic sky catalogues.

\end{abstract}


\section{Introduction}
\label{sec:intro}

The estimation of the redshift of distant astronomical sources (mainly galaxies and Active Galactic Nuclei, AGN) is a crucial part of modern cosmology \citep{2018MNRAS.478..592H} and extragalactic science \citep{2015ApJ...804..104M}. However with increasingly large data sets, in the era of high-precision cosmology, the requirements on the quality of galaxy redshift estimation can be very high \citep{2021PhRvD.103b3524M}. 
If, for example, in the instance of cosmological inference (say 3x2pt analysis, \citealp{2021OJAp....4E..13Z}), 
the redshifts were systematically a few percent higher or lower than the true redshifts (unknown to us and not included in the modelling), there could be a risk of inferring the incorrect cosmological model. Furthermore, redshifts are critical for extragalactic science, including galaxy formation and evolution, since they provide the third dimension and  time evolution (see \citealp{2017MNRAS.466..228E} and \citealp{2000AJ....120.2206F}). Their use in a variety of science disciplines, therefore leads to a strong need to understand their accuracy and precision. 

There are two main observational approaches to estimating redshifts, both of which have advantages and disadvantages; spectroscopy and photometry \citep{2001ApJS..135...41F}. Spectroscopic redshifts (`spec-$z$') are measured by identifying an emission/absorption feature in a galaxy's spectrum and comparing it to the known rest frame wavelength. Spec-$z$ estimations typically provide highly accurate redshift values, but can be expensive in terms of telescope time and are thus limited by the sample size, and also typically are more challenging to obtain for high redshift and low luminosity sources. 

By contrast, photometric redshifts (`photo-$z$') make use of photometry. When a spectrum is redshifted, spectral features move in and out of different photometric bands, giving changing measured magnitudes; thus, the photo-$z$ technique relies on the capacity to isolate the wavelength position of redshifted continuum features (e.g. Balmer or Lyman breaks). Hence instead of having a spectrum, we have a certain number of discrete photometric bands which must be mapped onto a redshift value. The need for multiple photometric bands is due to the redshift degeneracies present when one colour corresponds to multiple redshift values (e.g. confusion between different breaks). Wide multiwavelength coverage is necessary for photo-$z$ surveys to limit this effect. The primary benefit of using photo-$z$ is the derivation of redshift measurements for a much larger number of sources detected in imaging surveys, typically to higher magnitudes and redshifts. These low-cost photo-z estimates, however, are typically much less precise than spec-z estimates.

For many existing and forthcoming galaxy surveys like the Dark Energy Survey \citep{2005astro.ph.10346T}, The Vera C. Rubin Observatory Legacy Survey of Space and Time \citep{2009arXiv0912.0201L}, and Euclid \citep{2011arXiv1110.3193L}, the majority of galaxy redshifts will be based on photometry.

Photo-$z$'s themselves have multiple different approaches to their calculation, including template-fitting, machine learning (ML) and hybrid techniques. In the case of template fitting, a series of galaxy templates are selected and a chi-square fitting-like approach is performed where they are shifted in order to see what fits the photometry best. Some examples of template fitting codes include HYPERZ \citep{2000A&A...363..476B}, LE PHARE \citep{2006A&A...457..841I} and EAZY  \citep{2008ApJ...686.1503B}.

The machine learning (ML) approach instead utilises a sample of galaxies with \textit{both} photometric and spectroscopic values as the training set of the algorithm. The machine learning code then `learns' how to map the photometric colour-magnitudes of the training data onto the redshift values.  \citet{2019NatAs...3..212S} discusses several of these approaches to accurately estimating photo-z's, while also commenting on the challenge in achieving high redshift precision in large-scale galaxy surveys. Hybrid approaches typically attempt to combine template fitting and machine methods (see \citealp{2018MNRAS.477.5177D} and \citealp{2020MNRAS.498.5498H}).

There are now several studies seeking to rigorously assess and compare photo-z performance. \citet{2020MNRAS.499.1587S} investigated twelve photo-$z$ algorithms, where the codes were tested in ideal training and test data scenarios with mock data created for the Rubin Dark Energy Science Collaboration (DESC). Similarly, in \citet{2020A&A...644A..31E}, thirteen photo-z methods using either template-fitting or machine learning techniques were examined on Euclid-like data, providing a detailed comparison of their metrics. The goal of these studies is to develop photo-$z$ techniques appropriate to each survey, and to understand in advance what systematic biases need to be modelled and mitigated against.

One important aspect of the machine learning approach is a dependence on a reliable spectroscopic sample. This is one of the primary sources of systematic bias affecting photo-z estimation with machine learning. We hence need to consider the \textit{representativity} of the training data, since commonly, the spectroscopic redshifts do not span the full colour-space that the target data set might  (see \citealp{2017MNRAS.468.4323B}). This can lead to reduced performance in parts of colour-magnitude space poorly represented in the training set. Beyond representativity, another source of systematic error is incorrectly labelled training data i.e. incorrect spectroscopic redshifts, normally as a result of emission line confusion. This can risk the machine learning algorithm incorrectly `claiming' it is giving good predictions, because it is giving photo-$z$ predictions that agree with the spectroscopy - but if the spec-$z$ are themselves inaccurate, then it is very difficult to evaluate the true performance.

The challenge addressed by this paper is to try and understand the impact on photo-$z$ estimation (and specifically on the calculated posterior PDFs) in the non-idealised scenario where i) the training and test data have dramatically different distributions in colour-magnitude space and ii) some fraction of the spectroscopic data is mislabeled. We use the machine learning code GPz, which has already been tested for LSST and Euclid-like scenarios  (\citealp{2020MNRAS.499.1587S} and \citealp{2020A&A...644A..31E}). A number of photo-$z$ metrics are evaluated for a range of degradations, for not just point estimates, but also photo-$z$ PDF compared with true redshift, and photo-$z$ PDF compared with true redshift PDF. In Section \ref{sec:data} we will discuss the data used in this work (and what degradations we applied to them), in Section \ref{sec:est} we will describe how we calculated our photo-$z$ estimates, in Section \ref{sec:metrics} we describe the metrics used, in Section \ref{sec:res} we show our results, we discuss them in Section \ref{sec:disc}, and finally we conclude in Section \ref{sec:concl}.
The code for this work is available on GitHub at \url{https://github.com/nataliastylianou/Photo-z}.

\vspace{2mm}
\section{Data generation}
\label{sec:data}

\subsection{The Buzzard Simulation}
The data set used in this work is a sample of 100,000 galaxies from the Buzzard Flock synthetic sky catalogues\footnote{Also used in \cite{2020MNRAS.499.1587S}} (\citealp{2019arXiv190102401D}), with redshifts in the range $0<z<2.3$ and photometry in the LSST ugrizy bands\footnote{In particular we use the sample from \url{https://github.com/jfcrenshaw/pzflow}}. The Buzzard catalogues are constructed by first adding galaxies onto a dark matter only N-body simulation (in such a way as to be consistent with known lower-redshift luminosity functions), and then `observing' them by imposing a realistic set of observational properties and systematics. We chose this data set in order to have a mock catalogue of sufficient size with observational properties similar to LSST, and did not need some of the more complex physics that other simulations and mock catalogues might capture (as we are specifically focussing on the impact of degradation of the spectroscopic training set, not other effects e.g. stellar contamination).

\subsection{Normalising Flows}

In this work we will at points compare \textit{predicted} redshift PDF to \textit{true} redshift PDF. Galaxies only have one redshift, but given a set of photometric observations (bands and depths), there is a PDF that would represent perfect extraction of redshift information. If the joint $N+1$ (where $N$ is the number of bands) dimensional redshift magnitude distribution $p(z,\boldsymbol{m})$ (where $z$ is the redshift, and $\boldsymbol{m}$ was the photometry) were known perfectly, then the `true' redshift PDF for a set of observed magnitudes would be $p(z|\boldsymbol{m})$, the best possible estimation of the redshift (even though individual galaxies only have a single value for redshift). In a realistic observational scenario we will not a priori know $p(z,\boldsymbol{m})$ - but in this paper we will test how well we can reconstruct $p(z|\boldsymbol{m})$ in a scenario where we do know the full distribution from the simulation.

The joint distribution $p(z,\boldsymbol{m})$ is closely linked to the redshift distribution for the whole population, $N(z)$ (which is required for some science goals). Firstly, if $p(z,\boldsymbol{m})$ is perfectly known, the univariate distribution for $z$ can be found with a simple marginalisation: $N(z)=\int p(z,\boldsymbol{m}) \mathrm{d}\boldsymbol{m}$. Secondly, given a set of galaxies with `true' redshift PDFs $p(z|\boldsymbol{m})$, and the corresponding population distribution of colour-magnitudes $p(\boldsymbol{m})$ (itself a function of the underlying luminosity functions and the observational properties of the survey), $N(z)$ can again be recovered by performing the relevant weighted integral (`stacking'): $N(z)=\int p(z|\boldsymbol{m}) p(\boldsymbol{m}) \mathrm{d}\boldsymbol{m}$. Finally, if we knew $p(z,\boldsymbol{m})$ perfectly, then we could treat it as a prior in redshift for an unseen galaxy before any magnitudes were observed (or if only some of the magnitudes were known). For more realistic observational cases where we are merely estimating the true PDF, and don't perfectly know $p(z,\boldsymbol{m})$, a set of more complex approaches have been developed to convert a set of PDF estimates back to $N(z)$ (e.g. see \citealp{Malz2021}, which discusses more rigorously when a simple stacking approach is appropriate, and when it is not).

In the case of a simulation, we perfectly know both a) the intrinsic joint distribution of magnitudes and redshifts for the galaxies, and b) the selection function, which is applied in the simulation to construct the mock sky catalogue. However, in the case of real observations, we do not know neither the intrinsic $p(z,\boldsymbol{m})$ distribution, which is often what we intend to measure, nor do we typically perfectly understand the selection effects, although we have usually have some knowledge of certain constraints (e.g. detection thresholds).

The sample of sources from the Buzzard Flock synthetic sky catalogues all have single redshift values. In order to construct true redshift PDFs to compare against, we use a Normalising Flow \citep{2015arXiv150505770J} to model the $p(z,\boldsymbol{m})$ of the Buzzard sample. Normalizing Flows are tools that use sequences of invertible mappings to convert simple distributions into more complex ones. From the resulting Normalising Flow we can sample galaxies with redshifts, galaxies with redshift PDFs, and we can even apply transformations that correspond to degradations to create new Normalising Flows, which we can also sample from. To further build a Normalising Flow we use \textit{pzflow}, which is a package that models normalising flows \citep{Crenshaw_2021_pzflow}. The approach used here uses code from the Redshift Assessment Infrastructure Layers code (RAIL\footnote{\url{https://github.com/LSSTDESC/RAIL}}) and borrows heavily in approach from the example flow in \textit{pzflow}\footnote{\url{https://github.com/jfcrenshaw/pzflow/blob/main/pzflow/examples/examples.py}}.

We use this Normalising Flow sample with photometry, redshifts, and true redshift PDFs as our `no-degradation' sample. This data acts as our training, validation and test data for the `no-degradation' case, and also the test data for when the algorithm is trained on the degraded data.

\subsection{More Realistically Complex Training Set Imperfections}
\label{sec:degr}

Ideally for an ML-based calculation of photo-$z$, the training set would consist of perfectly redshift labelled sources with the exact same colour-magnitude distribution as the test data. This is in practice never achieved.

This paper will focus on two sources of training set imperfections with the aid of two degraders (taken from RAIL\footnote{\url{https://github.com/LSSTDESC/RAIL/tree/master/rail/creation/degradation}}), Inverse Redshift Incompleteness, which introduces sample incompleteness (i.e. where the training set is not representative of the test data) and the Emission Line Confusion, which includes spectroscopic systematics (i.e. training spectroscopic data are labelled with incorrect redshifts).

Other training set imperfections not discussed in this paper include AGN variability - consider that if the magnitudes of sources are changing over time then photometric redshift estimates risk becoming more unreliable (as the source redshift does not change) \citep{2015A&A...584A.106S}. Similarly, the blending of sources, and dust reddening depending on galactic coordinates, all result in inaccurate photometric redshift estimates \citep{2000ApJ...533..682C}.

\subsubsection{Inverse Redshift Incompleteness}

The Inverse Redshift Incompleteness Degrader attempts to replicate redshift incompleteness by applying a selection function inversely proportional to redshift. Its selection function probability is described by:
\begin{equation}
    p(z)=\min \left(1, \frac{z_{p}}{z}\right)
\end{equation}
where the $z_{p}$ term defines the pivot redshift, specifying the redshift where the incompleteness begins.

\begin{figure}[!ht]
    \centering
    \includegraphics[scale=0.65]{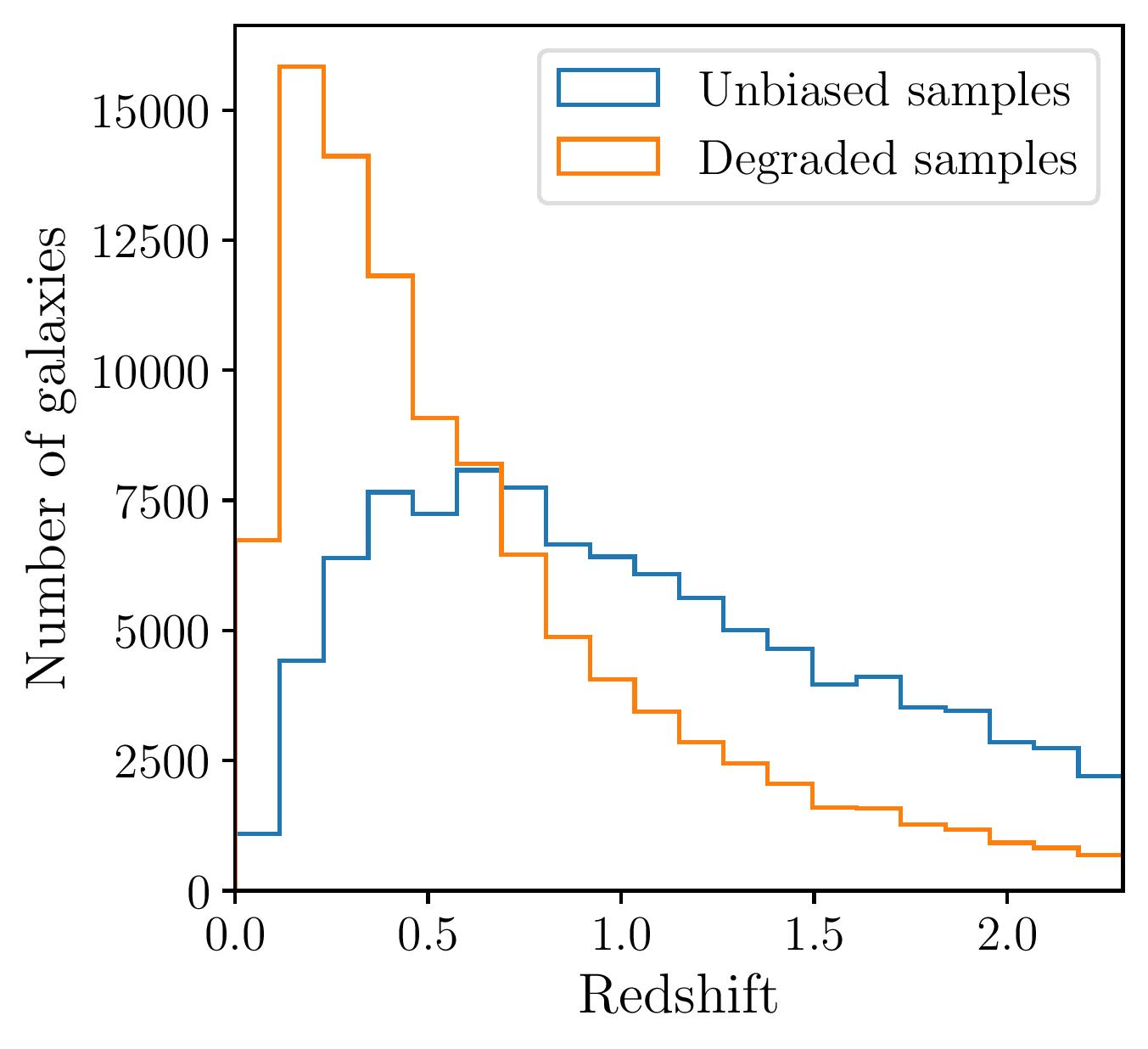}
    \caption{Redshift distributions for degraded and non-degraded (unbiased) data sets}
    \label{fig:Redshift_Incompleteness}
\end{figure}

Figure \ref{fig:Redshift_Incompleteness} shows the two different redshift distributions for the unbiased representative sample set and the degraded sample set through the Inverse Redshift Incompleteness Degrader. We set the pivot redshift for this degradation equal to $z_{p}=0.10$. As seen in Figure \ref{fig:Redshift_Incompleteness}, the degraded data set has most of its galaxies concentrated in the lower redshifts and very few lying in the high-redshift area. Such an effect can be caused by the difficulty in making spectroscopic measurements for high-redshift galaxies with faint photometric magnitudes (although of course in general low-luminosity sources at low-redshift may also be affected).

\subsubsection{Emission Line Confusion}

The Emission Line Confusion Degrader mimics the effect of spectroscopic systematic errors by simulating the confusion of different emission lines. Specifically, we used the Emission Line Confusion degrader to misidentify between 0.2\% and 10\% of OII lines as H$\alpha$ lines and vice versa (for a discussion of plausible line-confusions and percentages see \citealp{2021A&A...647A.117E}). Since OII emission lines have a wavelength of 3727\AA and H$\alpha$ lines a wavelength of 6563\AA, this confusion would consequently result in a larger spectroscopic redshift and the opposite misidentification would result in a smaller spectroscopic redshift.
\citet{2021A&A...647A.117E} has considered more complex models of emission line confusion, such as the misclassification of H$\alpha$ as OII lines for redshifts lower than 0.5 and between 1.4 and 2, OII lines as H$\alpha$ or Ly$\alpha$ lines for redshifts between 0.5 and 1.4, and for redshifts above 2 then Ly$\alpha$ lines are misidentified as OII lines. For the purposes of this paper, we proceeded to only use the OII and H$\alpha$ line confusion since further complexity would increase the realism by a small amount, as which line confusions are key will depend heavily on what spectroscopic training set is actually used (what resolution the spectrographs have, what lines were used for redshift measurement, what flag tolerance was used etc.).

\begin{figure}[!ht]
    \centering
    \includegraphics[scale=0.65]{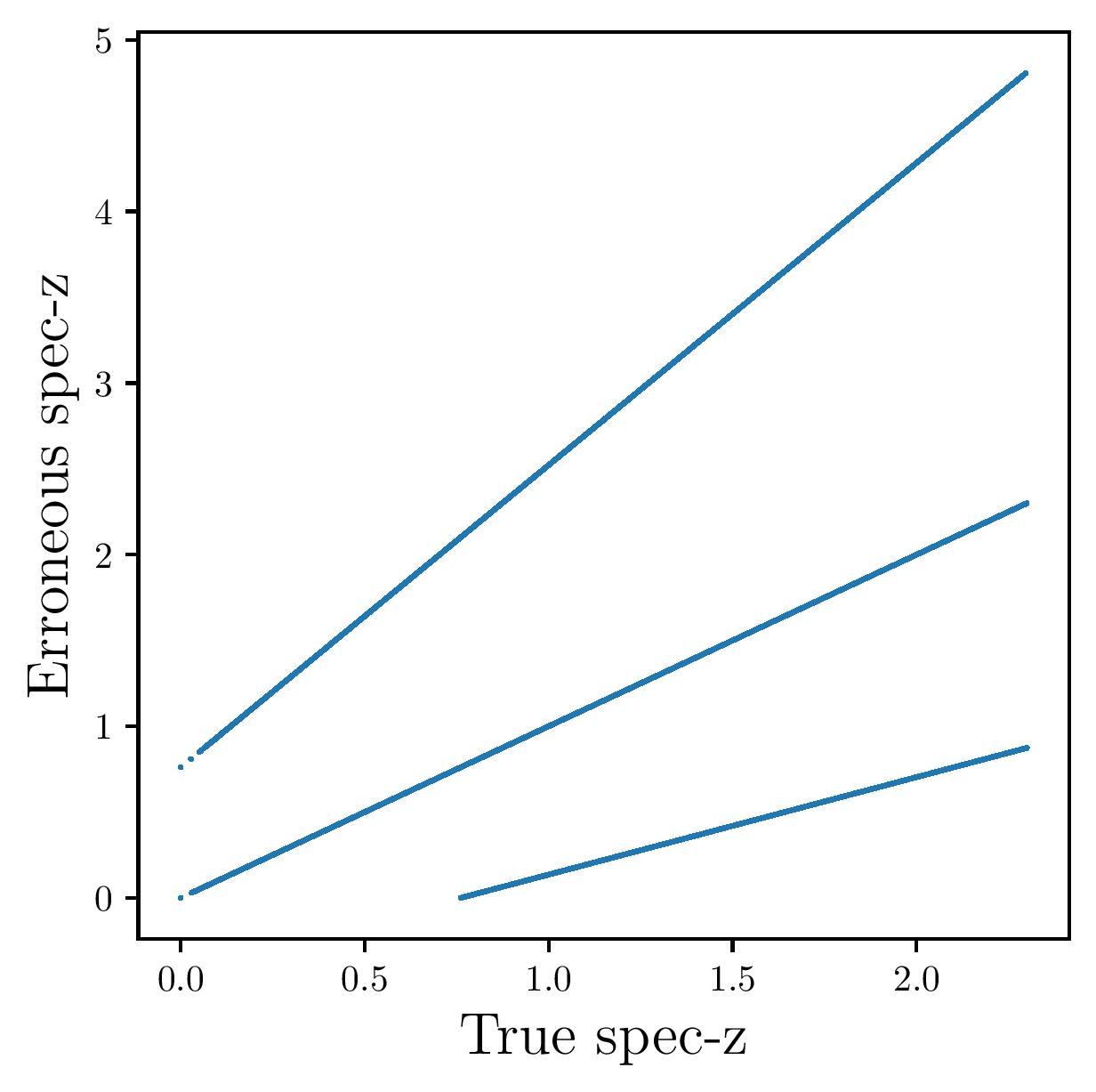}
    \caption{Degraded spectroscopic redshifts against non-degraded true redshifts data sets for the Buzzard photometry used in this study}
    \label{fig:Line_Confusion}
\end{figure}

Figure \ref{fig:Line_Confusion} shows the degraded spec-$z$'s with the emission line errors against the true non-degraded redshifts. The percentage of degradation used in Figure \ref{fig:Line_Confusion} is 5\% (which we will refer to as a `badness' parameter of 0.05). The central line (along the diagonal representing equality) in the plot represents where the true spec-$z$'s and the degraded spec-$z$'s are equal and there was no confusion in the line identification. Conversely, the two lines diverging from the one-to-one line illustrate where the OII and H$\alpha$ have been misclassified, resulting in a larger or smaller spec-z value. The more degradation the spec-z data endure, the more prominent the two diverging lines would be and the weaker the identity line would appear.

\section{Photo-$z$ estimation}
\label{sec:est}

The machine learning code we use in this work to estimate photo-zs is called GPz. It is a sparse Gaussian process code described in \citet{2016MNRAS.462..726A} and \citet{2016MNRAS.455.2387A}. GPz produces a point estimated mean and a variance that incorporates both the uncertainty due to intrinsic output noise, as well as due to low data density. Hence it accounts for training data insufficiency and galaxy magnitude degeneracies. GPz has been observed to be a fast, high performing machine learning code that typicically performs well for a range of metrics, normally particularly bias metrics. One weakness is that it outputs only point estimates with uncertainty as opposed to more general PDFs (uni-modal Gaussians instead of multi-modal PDFs), which can prove problematic for certain sources.

Ultimately, as with analogous ML-based codes, the performance of GPz is dependent on the spectroscopic training data used. As mentioned in Section 1, \citet{2020MNRAS.499.1587S} and \citet{2020A&A...644A..31E} tested GPz in the context of a photo-$z$ data challenge. In addition to this, GPz has been used in a number of separate studies, including \citet{2018MNRAS.475..331G} who tested inclusion of source size information, and \citet{2020MNRAS.498.5498H} who combined Gaussian mixture models (GMMs) with GPz to improve redshift estimation and ultimately accelerate photo-$z$ computation.

\section{Evaluation metrics}
\label{sec:metrics}
To assess the quality of photometric redshift estimations, there are a large number of different metrics that characterise in different ways how successful the estimates have been at predicting the true redshift. Metrics might compare just the photo-$z$ point estimate to the true redshift, the photo-$z$ PDF to the true redshift, or ultimately the photo-$z$ PDF to the true-PDF.

\subsection{Point estimate metrics}

We consider the following three metrics (see Section 6.3 of \citealp{2016MNRAS.462..726A}) to evaluate the point estimated outputs compared to single-valued true redshifts; these are i) the root mean squared error (RMSE), ii) the fraction retained for 15\% (FR15) and iii) the Bias. 

Specifically, the RMSE is defined as:
\begin{equation}
 \mathrm{RMSE}=\sqrt { \frac{1}{n} \sum_{1}^{n} \left(\frac{z_{spec}-z_{photo}}{1+z_{spec}} \right)^2 } 
\end{equation}

The FR15 classifies the fraction of catastrophic outliers with 15\% threshold defined as:
\begin{equation}
\mathrm{FR}15=\frac{100}{n} \Bigg\{ \Bigg|\frac{z_{spec}-z_{photo}}{1+z_{spec}}\Bigg| < 0.15\Bigg\}
\end{equation}

Finally, the Bias demonstrates how the photometric redshift deviates systematically from the true redshift.
\begin{equation}
 \mathrm{Bias}=\frac{1}{n} \sum_{1}^{n} \frac{z_{spec}-z_{photo}}{1+z_{spec}}
\end{equation}

All of these quantities can be calculated for the population as a whole, or considered as a function of redshift (or some other parameter).

\subsection{PDF metrics relative to true redshifts}

Beyond assessing the quality of point-estimates of the redshift, we might also wish to assess the quality of our uncertainty estimates, and the realism of the PDFs. We can hence consider i) the Probability Integral Transform (PIT) and ii) the Conditional Density Estimate (CDE) loss.

There are a broad range of other metrics in the literature to assess the quality of regression analyses that could be used to characterise the relationship between photo-$z$ PDFs and the true redshifts. One possibility is the coefficient of determination ($R^2$, \citealp{Wright1921}) metric, which characterises how much of the variation in the data is captured by a predictive model (where $R^2=1$ corresponds to perfect predictive power, $R^2=0$ would correspond to predicting the population mean each time, and $R^2<0$ corresponds to predictions poorer than simply guessing the population mean for each data point). This is in contrast to RMSE measurements for example, which can give comparisons of quality between two regressions, but do not solely by themselves give an indication of whether the model is capturing the full variance present in the data. $R^2$ has been used in astronomy for a range of applications, for example modelling light curves (\citealp{Shoji2020}) and testing goodness of fit in studies of magnetohydrodynamical turbulence (\citealp{Gonzalez-Casanova2018}). In terms of comparing PDFs rather than point estimates, $R^2$ was generalised to the Bayesian context in \citet{doi:10.1080/00031305.2018.1549100}. The $R^2$ metric captures several aspects of the quality of fit well, but can give misleading conclusions in others (see for example \citealp{Lewis-beck1990} for a discussion). We decided not to use $R^2$ in this work, as we felt the RMSE and PIT metrics together quantified the same behaviour $R^2$ captures (RMSE describing quality of fit, and PIT characterising how much of the true variance was captured by the calculated uncertainties). For example, when uncertainties on predictions are too small, the histogram of the PIT values shows a characteristic peak at 0 and 1, a peak at 0.5 when uncertainties are too large, and perfectly calibrated photo-$z$ PDFs give a uniform distribution (see Section \ref{sec:PIT}). In addition, the PIT distribution has been used to quantify the performance of photo-z PDF methods in many prior instances (\citealp{2018PASJ...70S...9T}, \citealp{2017MNRAS.468.4556F} and \citealp{2016arXiv160808016P}), making it more straightforwards to compare our work to other studies. However using (Bayesian) $R^2$ to compare photometric redshift PDFs would be an interesting study for future work.

\subsubsection{Probability Integral Transform (PIT)} \label{sec:PIT}

The Probability Integral Transform (PIT) metric seeks to assess the `realism' of PDFs for estimates for a population. For each prediction it takes the integral of individual PDFs from zero up to the true redshift, and then plots the distribution (a histogram) of those values. A histogram of PIT values is commonly used to assess how `realistic' a population of photo-$z$ PDFs is compared with the true redshifts. Ideally, the histogram would appear as a uniform distribution, which corresponds to the PDFs being perfectly calibrated. 

Although the PIT distribution is ideally flat, it is expected to be less flat for the biased training data set than for the representative data set. We can hence use this to see the impact of degradation on the predictions. The test statistics of the PIT distribution tell us about the deviation from that ideal flat distribution and are expected to be more discrepant for more biased training sets.

Outliers at high-redshifts often have underestimated means with large variances, but at low-redshifts they typically have overestimated means with small variances. There are more of the latter, so we expect an overabundance of low PIT values.

Note that the PIT is assessing the realism of the PDFs, not their information content per se;  \citet{2020MNRAS.499.1587S} showed that uninformative PDFs could get high PIT scores because they were very well calibrated PDFs, but did not give any information.

\subsubsection{Conditional Density Estimate (CDE) loss}

The Conditional Density Estimate (CDE) loss is the discrepancy of the ensemble of PDFs relative to true redshift-photometry distribution. 
See Section 4.2 in \citet{2020MNRAS.499.1587S} (also \citealp{2020A&C....3000362D}) for a full description and more detailed definition, but it is essentially the root-mean-square-error of the difference between the true and the predicted PDFs. However, in the absence of knowledge of the true PDF, it can still be determined up to a constant, even in the absence of knowledge
of the true underlying redshift-photometry distribution. 

The CDE loss metric essentially approximates the true posterior PDF from the estimated posterior PDF evaluated at the true redshift.  Therefore, the lower the CDE loss is, the better the predictions. The CDE loss might typically be expected to get worse the more biased and degraded the training set data is.

\subsubsection{Summary Statistics}

The exact shapes of the PIT  curves contain information about which way the estimated PDFs are biased. However, these curves can be summarised further if a single number is required. 

A Kolmogorov-Smirnov statistic test (KS test) can be used to find the maximum difference between the true and estimated cumulative distributions of the PIT values. The KS test outputs values from zero to one and the closer the KS value is to zero, the more uniform the PIT distribution is. Therefore, the KS values of the representative data sets are expected to be lower than for the biased data sets.

Similarly, a variant of the KS test is the Cramer-von Mises test (CvM test) which represents the mean-square difference between the cummulative distribution functions of estimated and true PDFs, and again would be expected to be lower for the representative training data sets.

Finally, a modification to the KS test is the Anderson-Darling test (AD test). The AD test describes the weighted mean-squared difference and gives more weight to discrepancies in the PIT distribution tails.

See \citet{2020MNRAS.499.1587S} and references within for the PIT, CDE loss and summary statistics tests.

\subsection{PDF metric relative to true PDFs}
Finally, comparison of the estimated and true photo-z posterior PDFs is possible, which we will do here with the Kullback Leibler Divergence metric. This metric has been used for photo-z PDF evaluation in \citet{2018AJ....156...35M}.

\subsubsection{Kullback Leibler Divergence (KLD)}
The Kullback Leibler Divergence (KLD) is the information content difference between the predicted PDF and the true PDF from which the data was generated (in our case via the Normalising Flow), and it is estimated for each PDF in the sample. Ideally the KLD value for each galaxy would be very small, signifying a low information loss from using estimated photo-$z$ PDFs instead of the true PDFs. 
\begin{equation}
\mathrm{KLD}=\int_{-\infty}^{\infty} \mathrm{PDF}_{\mathrm{true}}(z)  \log \left(\frac{\mathrm{PDF}_{\mathrm{true}}(z)}{\mathrm{PDF}_{\mathrm{estimated}}(z)} \right) dz
\end{equation}

\section{Results}
\label{sec:res}
To estimate photo-$z$'s with GPz, samples of 100,000 sources from the NF (either the degraded or non-degraded NF, as appropriate) are created. These data sets are then split into 3 subsequent sets. 20\% of the data was used for the training of the algorithm, 20\% for the validation process and the remaining 60\% of the data was used for the testing. For the non-degraded case, GPz is trained and validated on non-degraded data, and tested on non-degraded data. However, for the degraded case, GPz is trained and validated on the degraded data, but is still tested on the non-degraded data (i.e. both non-degraded and degraded predictors are applied to the same data). This allows us to probe the impact of training data imperfections on the quality of the predictor.

\begin{figure}[!ht]
    \centering
    \includegraphics[scale=0.55]{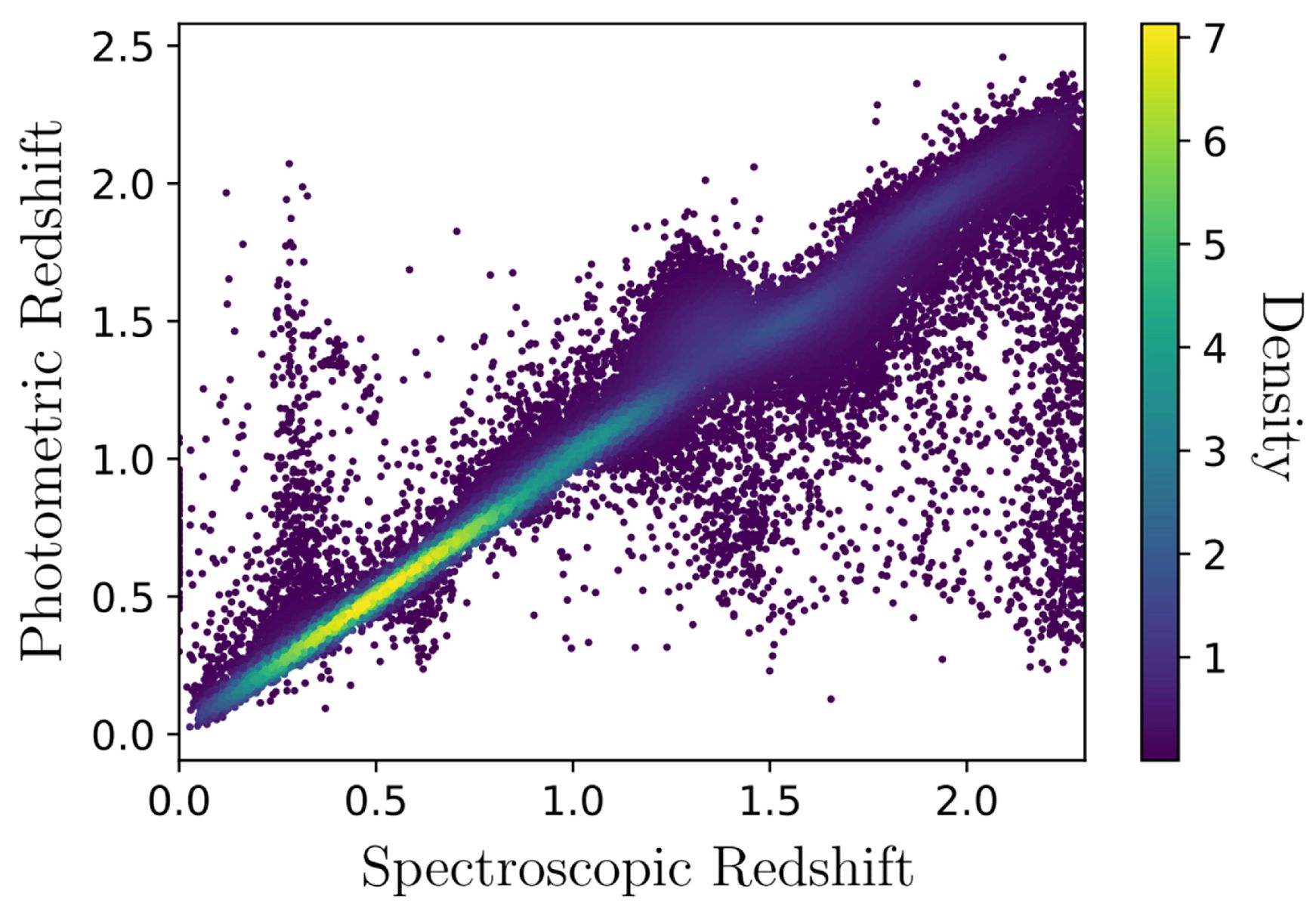}
    \caption{Photo-z predictions with the NF data}
    \label{fig:photoz_predictions_NF}
\end{figure}

The GPz performance with the NF data under no degradation is illustrated in Figure \ref{fig:photoz_predictions_NF}. We demonstrate the estimated photo-$z$'s in contrast to the spec-$z$'s, coloured by number density. It is evident that most galaxies have accurate photo-$z$'s consistent with their spec-$z$'s (lying on the diagonal), especially for $z<1$.

In the following sub-sections we describe the deviations from the best prediction performance quantitatively via the metrics previously discussed, as degradations are introduced.    

\subsection{Point Estimates}

\begin{figure*}[ht]
    \centering
    \begin{subfigure}[b]{0.4\textwidth}
        \centering
        \includegraphics[scale=0.69]{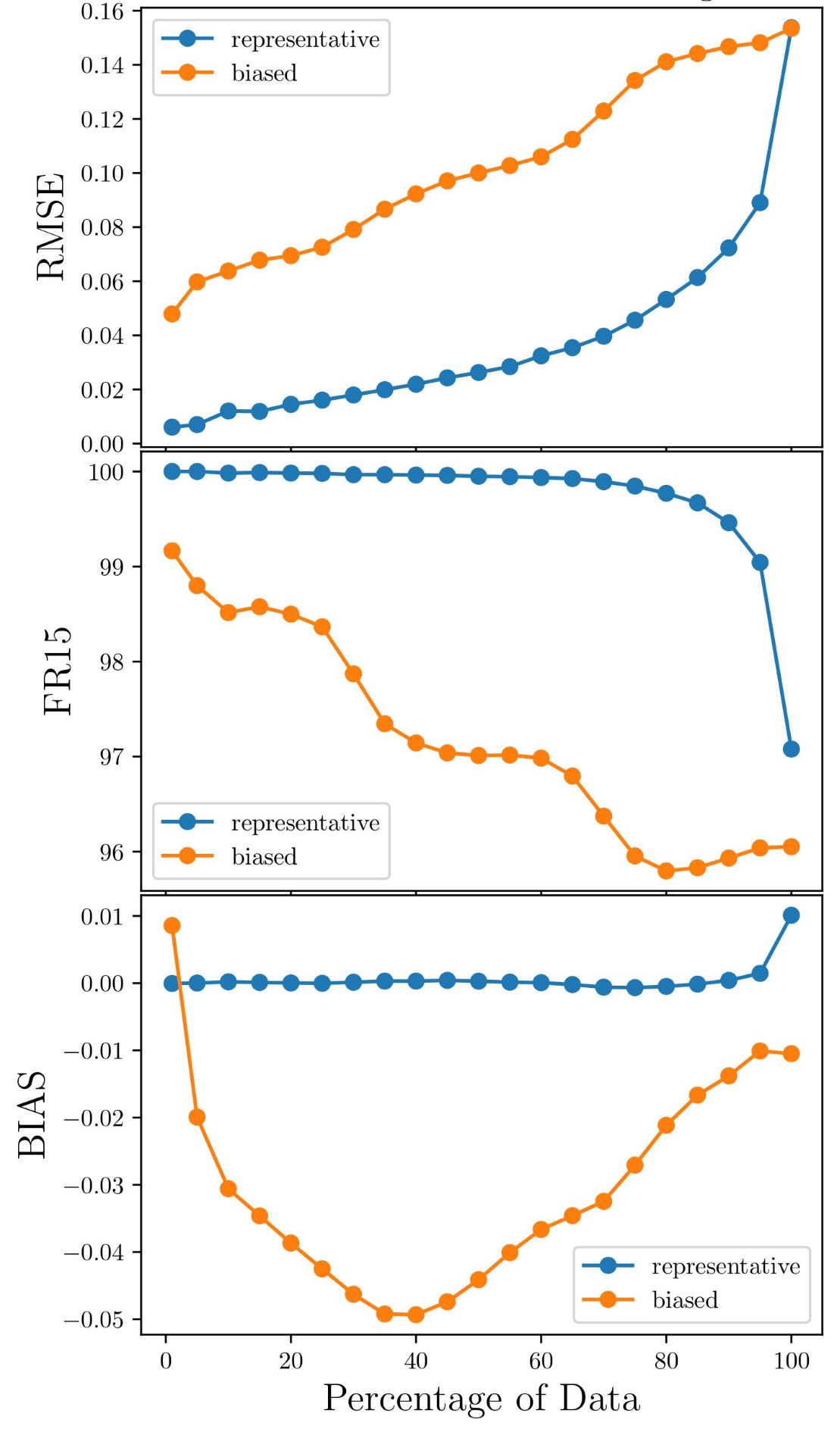}
        \caption{Emission Line Confusion Degrader}
        \label{fig:point_estimates_badness}
    \end{subfigure}
    \hfill
    \begin{subfigure}[b]{0.4\textwidth}
        \centering
        \includegraphics[scale=0.69]{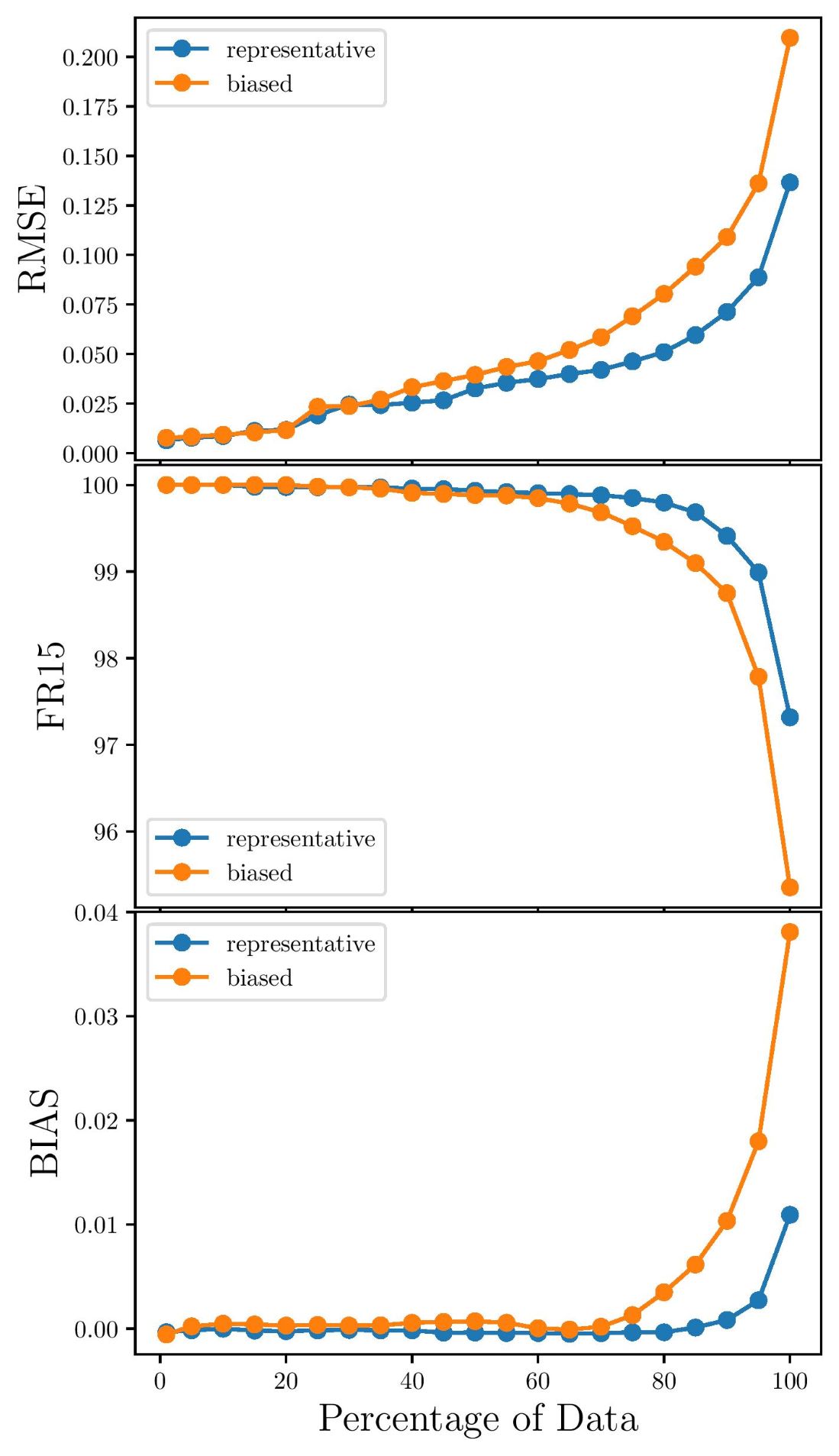}
        \caption{Inverse Redshift Incompleteness Degrader}
        \label{fig:point_estimates_z-pivot}
    \end{subfigure}
    \caption{Plots showing the RMSE, FR15 and BIAS metrics in terms of the Percentage of Data (galaxies ranked by uncertainty on prediction, with 0 being the lowest uncertainty) for the representative and biased samples of the two degraders}
    \label{fig:2_point_estimates_plots}
\end{figure*}

Figure \ref{fig:2_point_estimates_plots} exhibits how the RMSE, the FR15 and the Bias metrics vary with representative and non-representative data for the two degraders. The plots correspond to a degradation of 5\% for the Emission Line Confusion Degrader and a pivot redshift of 0.92 for the Inverse Redshift Incompleteness Degrader.
The RMSE of the biased data is larger than for the representative data in both degraders. Similarly, the biased data shows a lower FR15 in the two degraders than for the representative data.

Finally, the non-degraded data has a very flat Bias close to zero in both cases, whereas the degraded data has biases that deviate from zero in the negative and positive direction respectively for each degrader.

Importantly however, throughout all of these metrics, the Emission Line Confusion degrader plots illustrate a significantly larger discrepancy between the representative and degraded data (at least for the experimental setup considered here).

\subsection{PDF relative to true redshift}

\begin{figure*}[!ht]
    \centering
    \begin{subfigure}[b]{0.38\textwidth}
        \centering
        \includegraphics[scale=0.46]{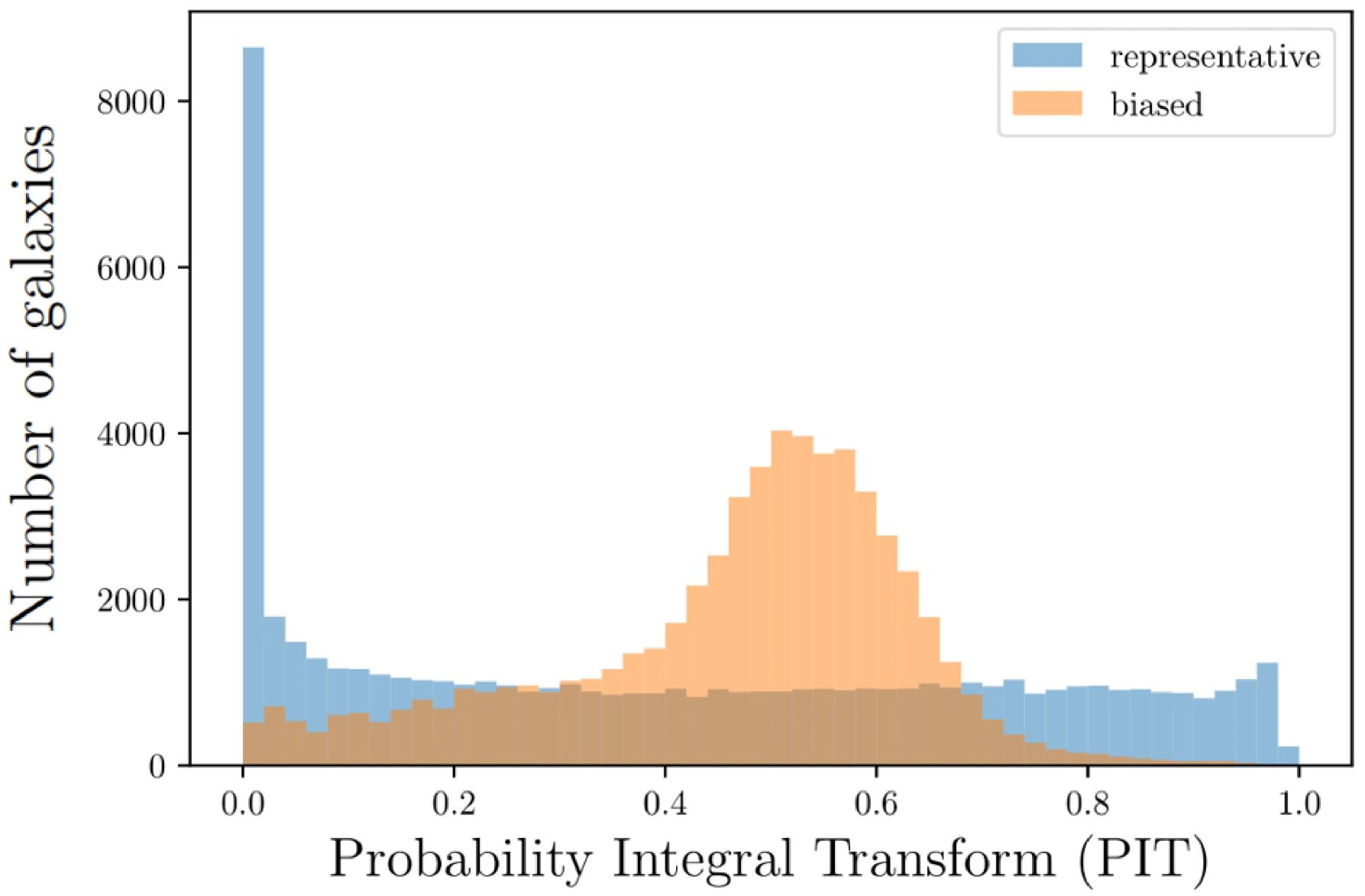}
        \caption{Emission Line Confusion Degrader}
        \label{fig:PIT_badness}
    \end{subfigure}%
    \hfill
    \begin{subfigure}[b]{0.38\textwidth}
        \centering
        \includegraphics[scale=0.46]{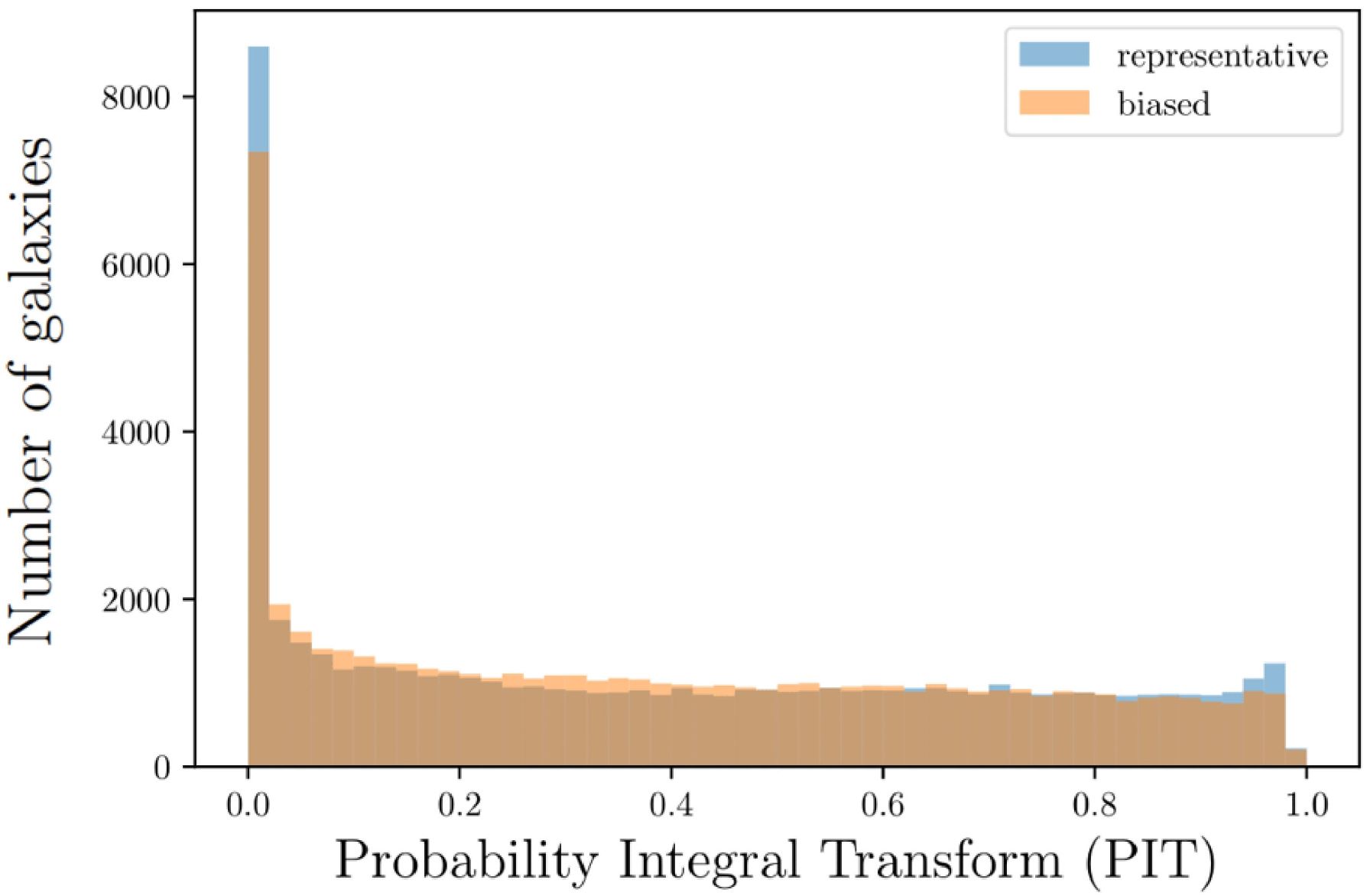}
        \caption{Inverse Redshift Incompleteness Degrader}
        \label{fig:PIT_z-pivot}
    \end{subfigure}%
    \caption{Histograms of PIT values for the representative and biased data sets of the two degraders}
    \label{fig:2_PIT_plots}
\end{figure*}

In Figure \ref{fig:2_PIT_plots} we show the two plots of the PIT distribution of degraded and non-degraded NF data for degradations of 0.05 badness and a 0.92 pivot redshift respectively. 
Both plots indicate a nearly flat distribution for the representative data with the exception of a spike in the very beginning and a considerably smaller spike at the very end of the distribution. The biased data in the Emission Line Confusion Degrader shows a peak at about 0.5, while for the Inverse Redshift Incompleteness Degrader the biased data has similar PIT values to the representative data.

\begin{figure*}[!ht]
    \centering
    \begin{subfigure}[b]{0.4\textwidth}
        \centering
        \includegraphics[scale=0.7]{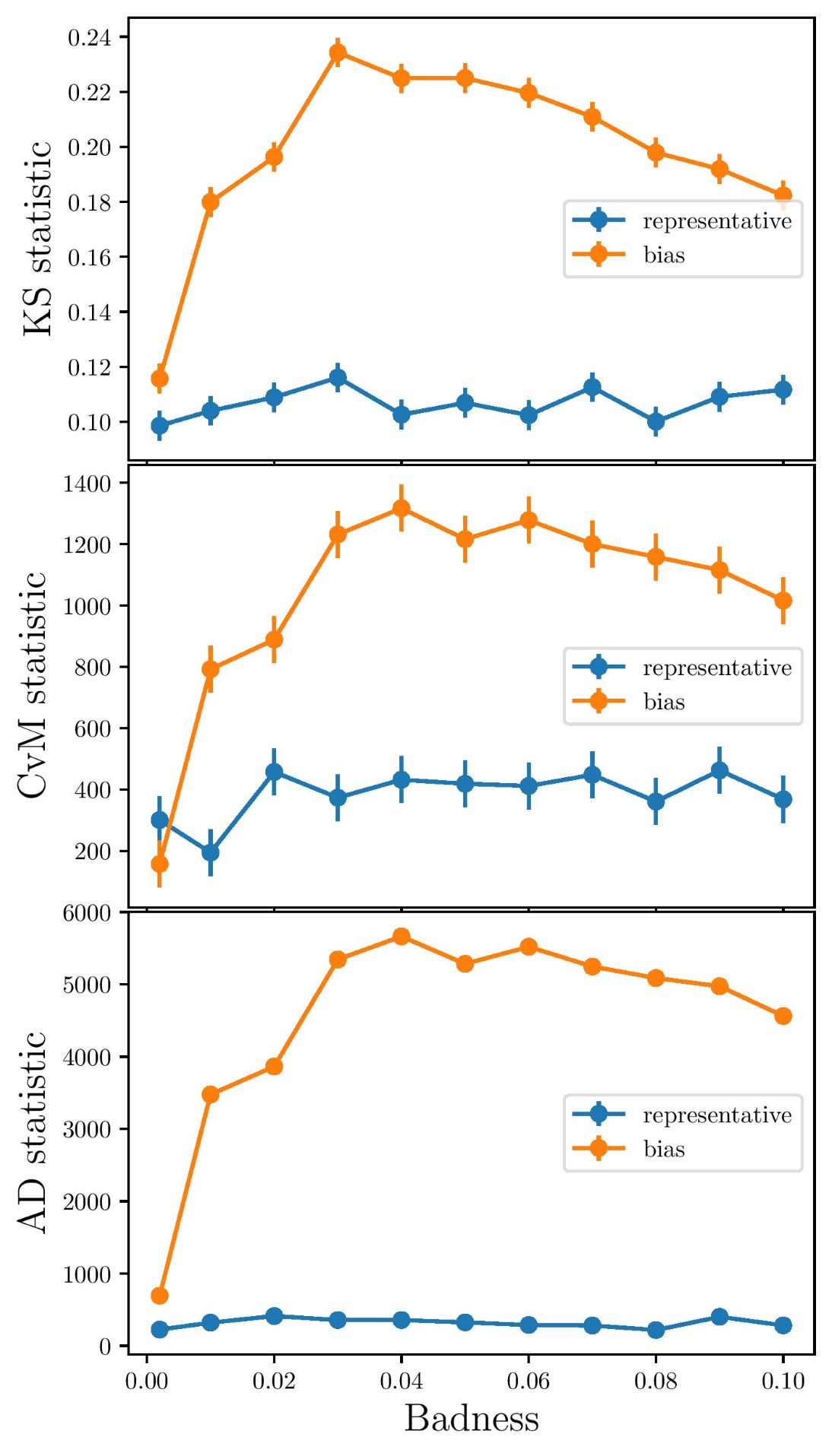}
        \caption{Emission Line Confusion Degrader}
        \label{fig:PIT_summary_statistics_badness}
    \end{subfigure}%
    \hfill
    \begin{subfigure}[b]{0.4\textwidth}
        \centering
        \includegraphics[scale=0.7]{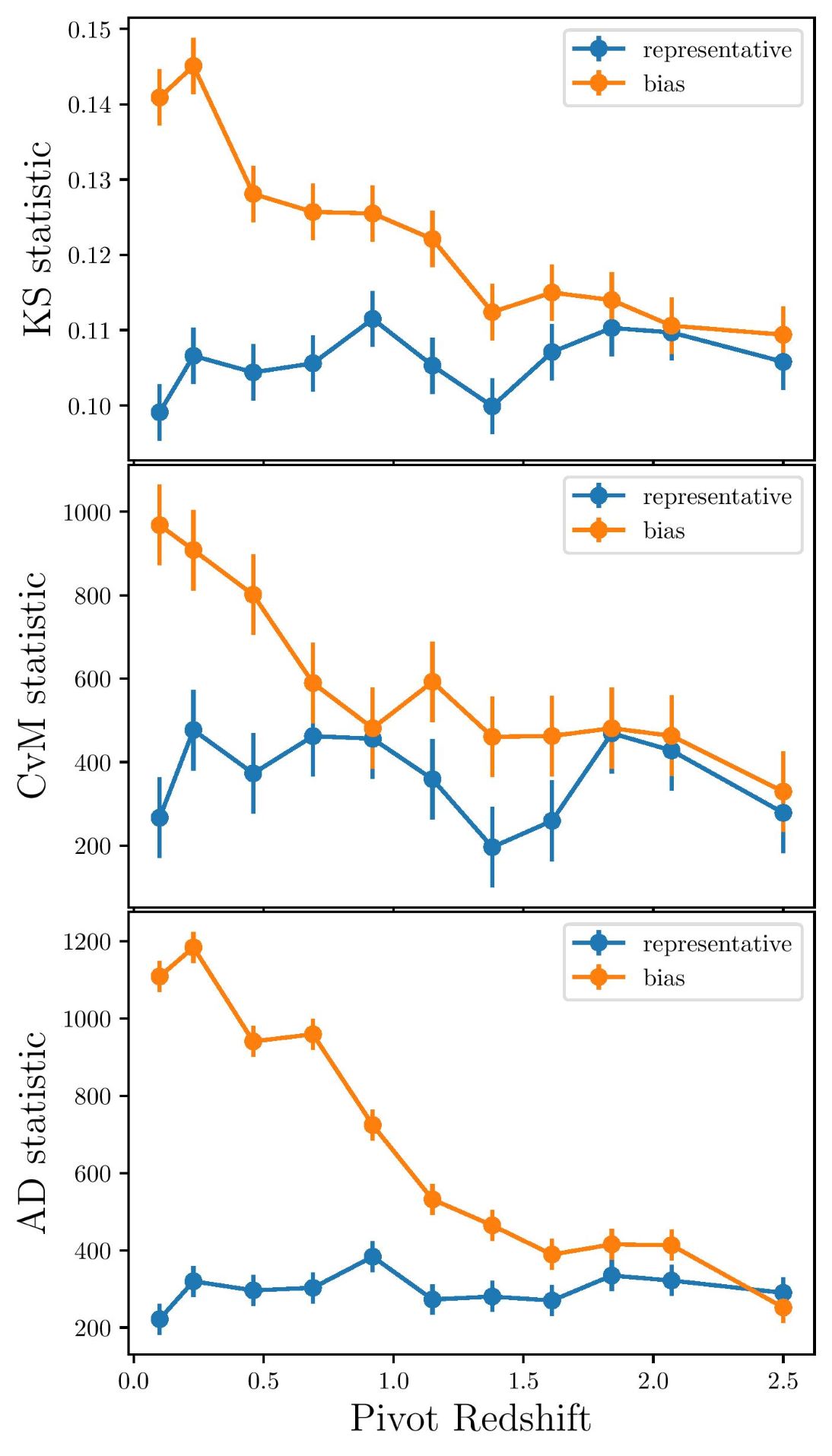}
        \caption{Inverse Redshift Incompleteness Degrader}
        \label{fig:PIT_summary_statistics_z-pivot}
    \end{subfigure}%
    \caption{Plots showing the PIT Summary Statistics for the two degraders}
    \label{fig:2_PIT_summary_statistics_plots}
\end{figure*}

The summary statistics of PIT (KS, CvM and AD) are shown in Figure \ref{fig:2_PIT_summary_statistics_plots} for different badness and pivot redshifts parameters. The badness parameter was varied from 0.002 to 0.10 (from best to worst) and the pivot redshift from 0.10 to 2.5 (from worst to best) to evaluate performance.
For both degraders, the metrics can be seen to be inconsistent with the non-degraded case (apart from for the extreme where no degradation is taking place). All of the representative data points are identical in every degradation scenario; therefore the standard deviation from their scatter was calculated and included as a representative error bar for all the summary statistics values\footnote{We could have run this whole analysis a large number of times to calculate error bars but this would have been highly computationally expensive.}.
In the Emission Line Confusion Degrader's case, the metrics get very rapidly poorer as the badness degradation parameter increases, and then flattens off for higher values. In the Inverse Redshift Incompleteness Degrader's case, there is a distinct rise in the values of the summary statistics for the lower pivot redshifts (more extreme degradation). 

\begin{figure*}[!ht]
    \centering
    \begin{subfigure}[b]{0.4\textwidth}
        \centering
        \includegraphics[scale=0.5]{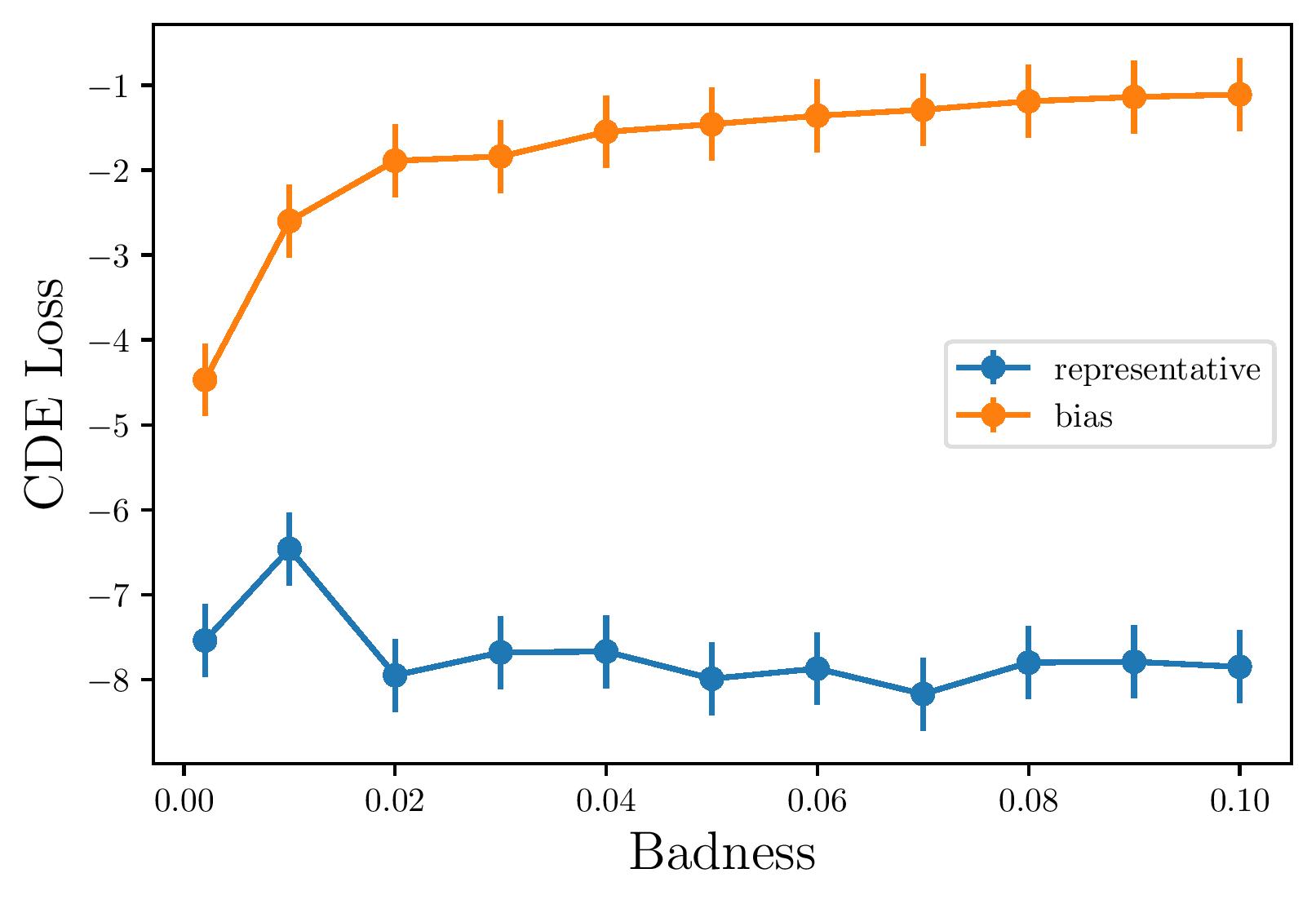}
        \caption{Emission Line Confusion Degrader}
        \label{fig:CDE loss vs Badness}
    \end{subfigure}%
    \hfill
    \begin{subfigure}[b]{0.4\textwidth}
        \centering
        \includegraphics[scale=0.5]{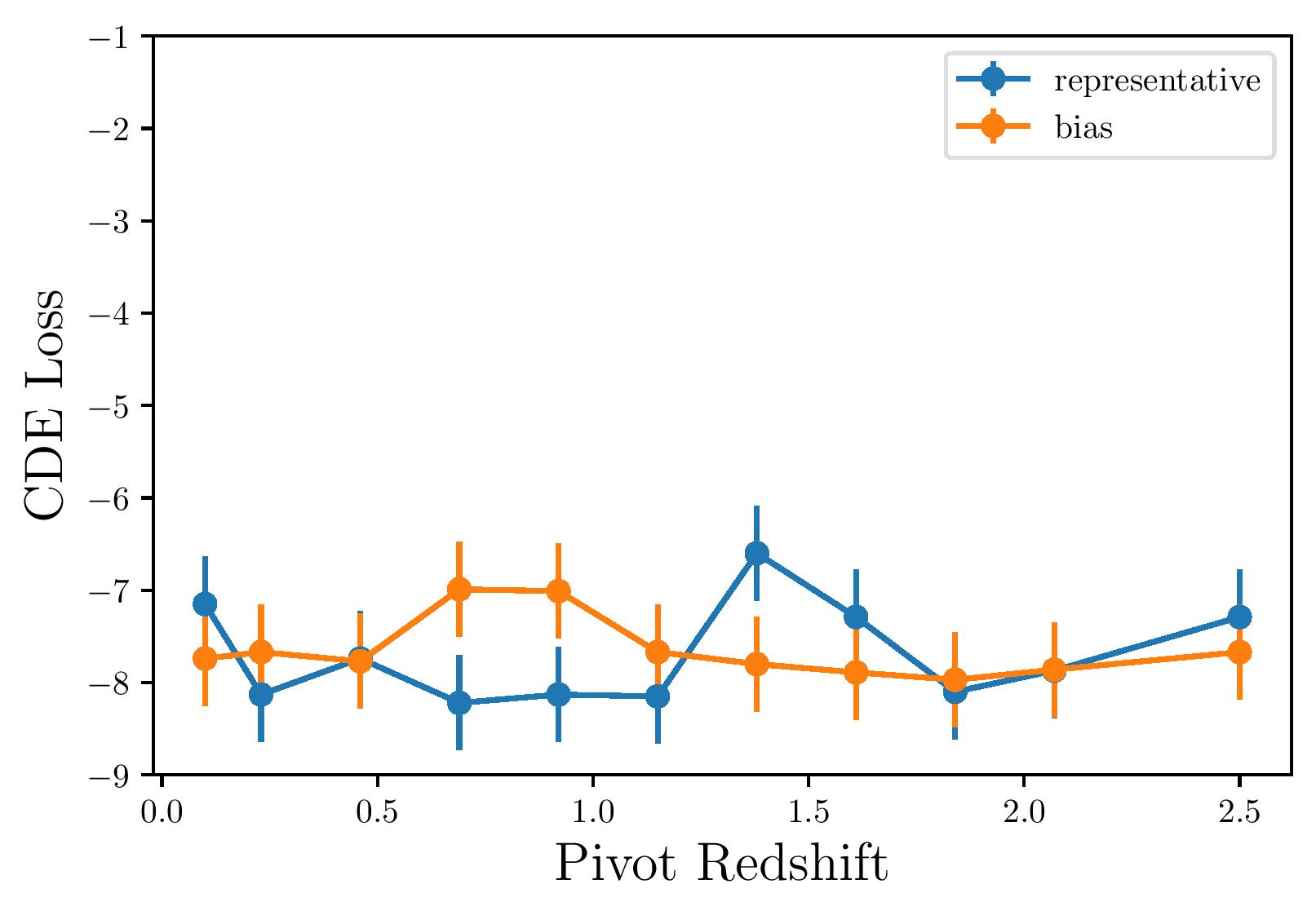}
        \caption{Inverse Redshift Incompleteness Degrader}
        \label{fig:CDE loss vs z-pivot}
    \end{subfigure}%
    \caption{Plots of the CDE loss against a gradient of degradations corresponding to the two degraders}
    \label{fig:2_CDE_loss_plots}
\end{figure*}

The CDE loss is shown as a function of badness and pivot redshift in Figure \ref{fig:2_CDE_loss_plots} with the standard deviation of the representative values as error bars and hence two different trends are observed. The CDE loss seems to in fact be very similar between the degraded and non-degraded predictions for the Inverse Redshift Incompleteness Degrader case. Conversely, for the Emission Line Confusion Degrader case, the CDE loss is low for the representative data and high for the biased data (even for very low fractions of line confusion).

\subsection{Estimated PDF relative to true PDF}

\begin{figure*}[!ht]
    \centering
    \begin{subfigure}[b]{0.4\textwidth}
        \centering
        \includegraphics[scale=0.45]{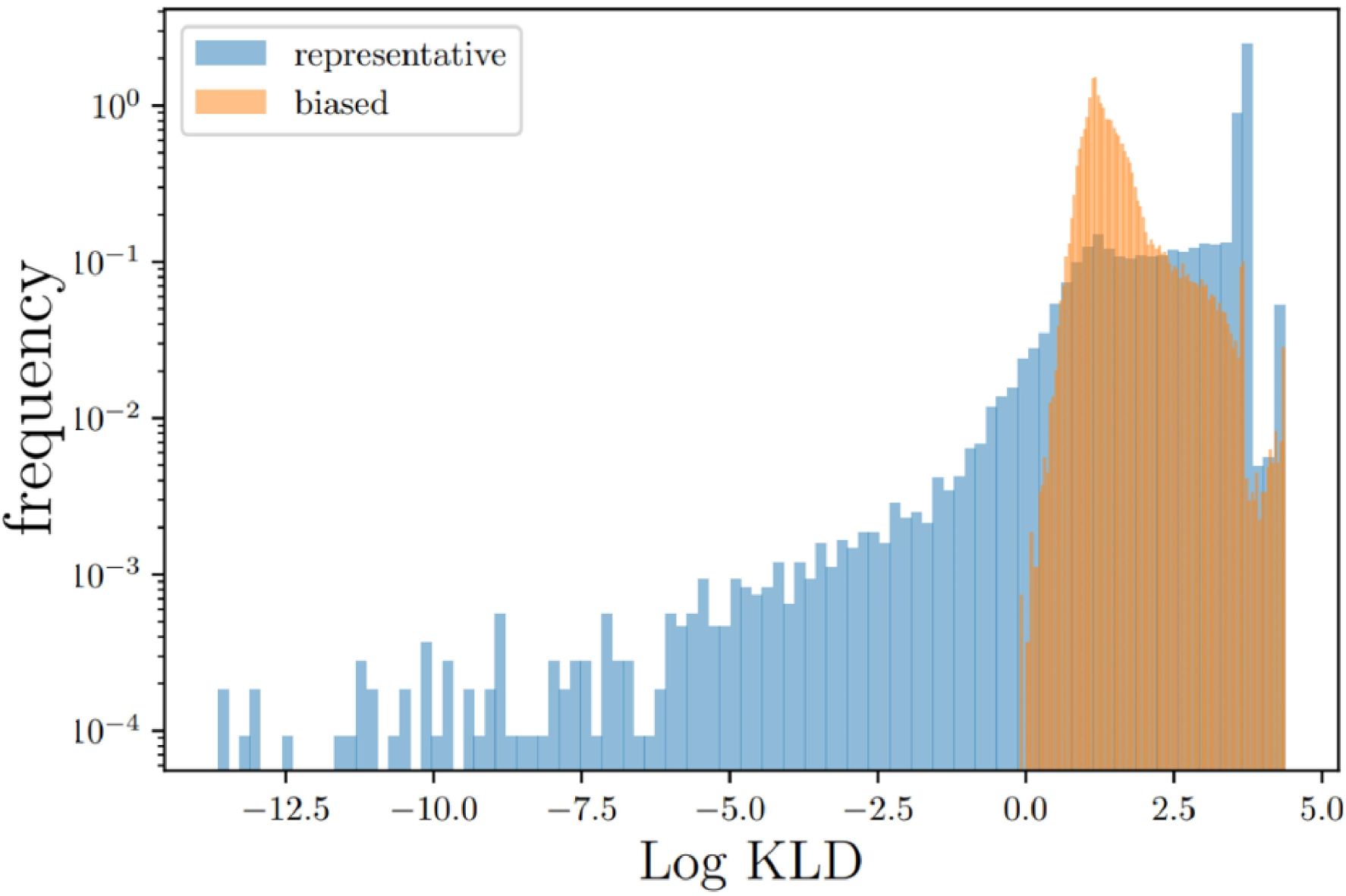}
        \caption{Emission Line Confusion Degrader}
        \label{fig:KLD_badness}
    \end{subfigure}%
    \hfill
    \begin{subfigure}[b]{0.4\textwidth}
        \centering
        \includegraphics[scale=0.45]{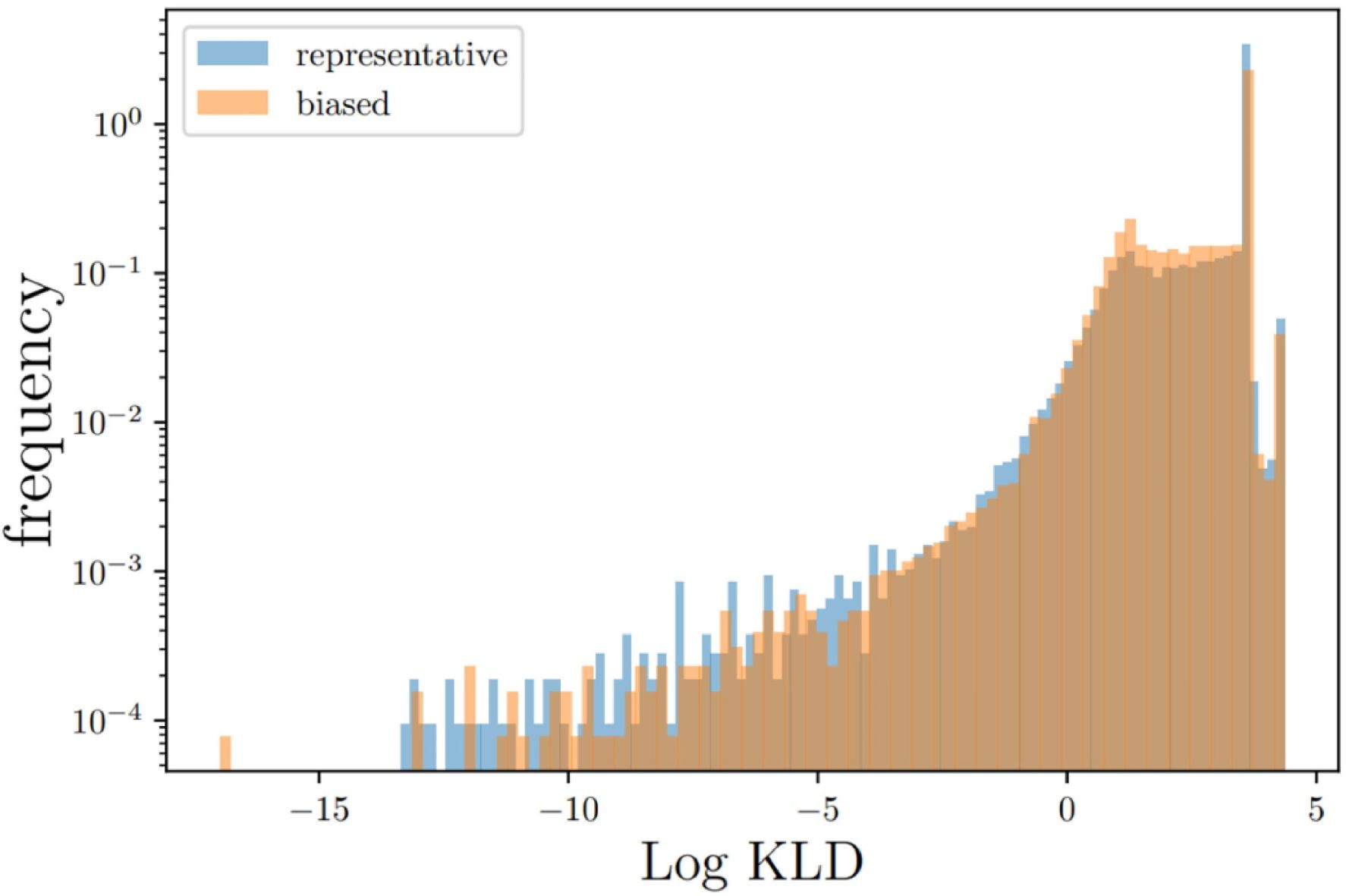}
        \caption{Inverse Redshift Incompleteness Degrader}
        \label{fig:KLD_z-pivot}
    \end{subfigure}%
    \caption{Plots showing the KLD metric for the two degraders}
    \label{fig:2_KLD_plots}
\end{figure*}

The logarithmic distribution of the KLD values for the two degraders is shown in Figure \ref{fig:2_KLD_plots}, for badness 0.05 and pivot redshift 0.92 respectively. The representative and bias data in the Inverse Redshift Incompleteness Degrader have comparable KLD distributions. However, in the Emission Line Confusion Degrader case, the representative data have a significant number of their KLD values below zero, unlike their corresponding biased data.

\vspace{5mm}
\section{Discussion}
\label{sec:disc}

The overall results for the two degraders exhibit similar behaviour, in that both of the degraded data show a decline in performance with increasing bias in the training set, while the non-degraded, representative data follow a consistent and generally good performance outcome for all metrics. It is also noted that the degradation of the Emission Line Confusion is generally more extreme than the Inverse Redshift Incompleteness degradation, and hence especially the point estimate plots of the former show greater discrepancies than of the latter.

The overabundance of low PIT values is found to be as expected (see Section \ref{sec:metrics}). The under-representation of high PIT values indicates that GPz is slightly too conservative with the variances (c.f. performance in the DC1 experiment, see \citealp{2020MNRAS.499.1587S}).

Based on the summary statistics and the CDE loss for the Emission Line Confusion, the metrics get worse quite fast from degradations of 0.2\% to about 3\% and then seem relatively flat from a degradation of 3\% onwards. This shows that there are only modest gains to be found from decreasing spec-z contamination fraction when above $\sim3\%$ - it appears it must be brought below $\sim1-2\%$ for the real improvements to show (at least for this experimental setup).

We note that for the Inverse Redshift Incompleteness case, we would potentially expect the biased set to have higher CDE loss for lower pivot redshifts, where the bias of the training set is stronger, and lower CDE loss at higher pivot redshift, where the distributions are very similar. This is not observed here - instead the bias and the representative data show a consistent performance outcome with low CDE loss throughout the training set degradations (although degradation did affect the other metrics). In other words, CDE loss appears not to be affected by redshift incompleteness. Similar behaviour is seen for the KLD metric; Emission Line confusion impacts this metric strongly, preventing Log KLD values below 0 being achieved for any galaxies, wheras for the incompleteness degradation the distribution of KLD values are comparable.

Regarding the performance sensitivity of GPz to the inverse redshift incompleteness, we can say that above a 1.5 pivot redshift the bias data point metrics and summary statistics are generally good for the representative data, as expected, but they get increasingly worse for pivot redshifts below 0.5 for the biased data. This redshift threshold is dependent on the data set used and hence it would likely be different if the original sample had a different redshift and colour-magnitude distribution. Nonetheless, for this and any comparable training set distributions, we would caution on sample incompleteness reaching below 0.5 redshift, as the impact on the relevant metrics then starts to become very substantial. 

Our results are in agreement with the findings of \citet{2014MNRAS.444..129C}, where the impact of incompleteness and incorrect spectroscopic redshifts was investigated using N-body-spectro-photometric simulations, although we study a broader range of metrics, including evaluations of PDF quality. They also found redshift incompleteness was potentially not as impactful as the emission line confusion in terms of impact on photo-z estimator performance. \citet{2014MNRAS.444..129C} demonstrated that incorrect redshifts have the most severe impact on the accuracy of photo-z estimators due to their significant degradation on the training set. In particular, they also found that 1\% is approximately the tolerable fraction for spectroscopic line confusion (before critically affecting cosmological biases), which is in accordance with our results.

\section{Conclusion}
\label{sec:concl}

To simulate imperfections in spectroscopic redshift training sets for photo-z's estimation we used two degraders to replicate emission line confusion and inverse redshift incompleteness. We compared photo-$z$ based on these biased data sets of increasing degradation with a set of representative data (drawn from the Buzzard Flock synthetic sky catalogues, constructed to be comprable to Rubin-LSST), and calculated a range of metrics that quantified how much poorer degradation of the training data made the resulting photo-$z$ estimates. 

It is clear that, broadly, the more biased the data are, and the larger the mismatch between the training and test set is, the worse the metrics and the overall performance of GPz (and likely any other ML-based photo-z estimator) is. Typically, the emission line confusion had a much greater impact on the metrics, with the CDE-loss and KLD metrics in particularly being only very weakly impacted by redshift incompleteness.

We have shown that for samples comparable to those used in this study, the incompleteness pivot redshift (for a sample spanning approximately $0<z<2.3$) should not reach below a redshift of 0.5, as this greatly affects the accuracy. Similarly, we have shown that the emission line confusion fraction may only be worth improving if it can be reduced below 1-2\%, since the decline in the metrics performance is dramatic after that, but relatively flat before. 

There are a large number of training set imperfection scenarios, each of which typically will affect photo-$z$ quality to a greater or lesser degree, depending on the exact properties of a specific survey. In this paper we have considered the sensitivity of machine learning based photometric redshift estimation under two training set imperfection scenarios, specifically looking towards the upcoming Rubin-LSST survey. Our results provide an insight into the level of tolerance of training set degradation needed for future large-scale studies, and how badly photo-$z$ predictions can be affected if not mitigated for.

\section*{Acknowledgments}
NS was supported by a 2021 LSSTC Enabling Science grant, as an “ISSC Ambassador”, a scheme to support student researchers working on projects connected to Rubin/LSST science. The program aims to build links between the Informatics and Statistics Science Collaboration (ISSC) and the other Rubin/LSST Science Collaborations (in this case the Dark Energy Collaboration, DESC).
This work was built upon a hack-tutorial at `Quarks to Cosmos with AI', a conference supported by the NSF AI Institute: Physics of the Future, NSF PHY-2020295.
AIM acknowledges support from the Max Planck Society and the Alexander von Humboldt Foundation in the framework of the Max Planck-Humboldt Research Award endowed by the Federal Ministry of Education and Research.
PH acknowledges generous support from the Hintze Family Charitable Foundation through the Oxford Hintze Centre for Astrophysical Surveys.
JFC is also an ISSC Ambassador under a 2021 LSSTC Enabling Science grant, and is supported by the U.S. Department of Energy, Office of Science, under Award DE-SC0011665, as well as the National Science Foundation, Division Of Astronomical Sciences, under Award AST-1715122. 
This work received software contribution offered from LIneA to DESC via the LSST international in-kind contribution program. 
The authors thank all developers of the RAIL package, which was key to the development of this project.

\software{
            pzflow,
            RAIL,
            GPz,
            cdetools,
            cde-diagnostics \citep{zhao2021diagnostics},
            qp,
          astropy \citep{2013A&A...558A..33A,2018AJ....156..123A}}

\bibliography{sample631,issc}{}

\begin{thebibliography}{}
\expandafter\ifx\csname natexlab\endcsname\relax\def\natexlab#1{#1}\fi
\providecommand{\url}[1]{\href{#1}{#1}}
\providecommand{\dodoi}[1]{doi:~\href{http://doi.org/#1}{\nolinkurl{#1}}}
\providecommand{\doeprint}[1]{\href{http://ascl.net/#1}{\nolinkurl{http://ascl.net/#1}}}
\providecommand{\doarXiv}[1]{\href{https://arxiv.org/abs/#1}{\nolinkurl{https://arxiv.org/abs/#1}}}

\bibitem[{{Almosallam} {et~al.}(2016{\natexlab{a}}){Almosallam}, {Jarvis}, \&
  {Roberts}}]{2016MNRAS.462..726A}
{Almosallam}, I.~A., {Jarvis}, M.~J., \& {Roberts}, S.~J. 2016{\natexlab{a}},
  \mnras, 462, 726, \dodoi{10.1093/mnras/stw1618}

\bibitem[{{Almosallam} {et~al.}(2016{\natexlab{b}}){Almosallam}, {Lindsay},
  {Jarvis}, \& {Roberts}}]{2016MNRAS.455.2387A}
{Almosallam}, I.~A., {Lindsay}, S.~N., {Jarvis}, M.~J., \& {Roberts}, S.~J.
  2016{\natexlab{b}}, \mnras, 455, 2387, \dodoi{10.1093/mnras/stv2425}

\bibitem[{{Astropy Collaboration} {et~al.}(2013){Astropy Collaboration},
  {Robitaille}, {Tollerud}, {Greenfield}, {Droettboom}, {Bray}, {Aldcroft},
  {Davis}, {Ginsburg}, {Price-Whelan}, {Kerzendorf}, {Conley}, {Crighton},
  {Barbary}, {Muna}, {Ferguson}, {Grollier}, {Parikh}, {Nair}, {Unther},
  {Deil}, {Woillez}, {Conseil}, {Kramer}, {Turner}, {Singer}, {Fox}, {Weaver},
  {Zabalza}, {Edwards}, {Azalee Bostroem}, {Burke}, {Casey}, {Crawford},
  {Dencheva}, {Ely}, {Jenness}, {Labrie}, {Lim}, {Pierfederici}, {Pontzen},
  {Ptak}, {Refsdal}, {Servillat}, \& {Streicher}}]{2013A&A...558A..33A}
{Astropy Collaboration}, {Robitaille}, T.~P., {Tollerud}, E.~J., {et~al.} 2013,
  \aap, 558, A33, \dodoi{10.1051/0004-6361/201322068}

\bibitem[{{Astropy Collaboration} {et~al.}(2018){Astropy Collaboration},
  {Price-Whelan}, {Sip{\H{o}}cz}, {G{\"u}nther}, {Lim}, {Crawford}, {Conseil},
  {Shupe}, {Craig}, {Dencheva}, {Ginsburg}, {VanderPlas}, {Bradley},
  {P{\'e}rez-Su{\'a}rez}, {de Val-Borro}, {Aldcroft}, {Cruz}, {Robitaille},
  {Tollerud}, {Ardelean}, {Babej}, {Bach}, {Bachetti}, {Bakanov}, {Bamford},
  {Barentsen}, {Barmby}, {Baumbach}, {Berry}, {Biscani}, {Boquien}, {Bostroem},
  {Bouma}, {Brammer}, {Bray}, {Breytenbach}, {Buddelmeijer}, {Burke},
  {Calderone}, {Cano Rodr{\'\i}guez}, {Cara}, {Cardoso}, {Cheedella}, {Copin},
  {Corrales}, {Crichton}, {D'Avella}, {Deil}, {Depagne}, {Dietrich}, {Donath},
  {Droettboom}, {Earl}, {Erben}, {Fabbro}, {Ferreira}, {Finethy}, {Fox},
  {Garrison}, {Gibbons}, {Goldstein}, {Gommers}, {Greco}, {Greenfield},
  {Groener}, {Grollier}, {Hagen}, {Hirst}, {Homeier}, {Horton}, {Hosseinzadeh},
  {Hu}, {Hunkeler}, {Ivezi{\'c}}, {Jain}, {Jenness}, {Kanarek}, {Kendrew},
  {Kern}, {Kerzendorf}, {Khvalko}, {King}, {Kirkby}, {Kulkarni}, {Kumar},
  {Lee}, {Lenz}, {Littlefair}, {Ma}, {Macleod}, {Mastropietro}, {McCully},
  {Montagnac}, {Morris}, {Mueller}, {Mumford}, {Muna}, {Murphy}, {Nelson},
  {Nguyen}, {Ninan}, {N{\"o}the}, {Ogaz}, {Oh}, {Parejko}, {Parley}, {Pascual},
  {Patil}, {Patil}, {Plunkett}, {Prochaska}, {Rastogi}, {Reddy Janga},
  {Sabater}, {Sakurikar}, {Seifert}, {Sherbert}, {Sherwood-Taylor}, {Shih},
  {Sick}, {Silbiger}, {Singanamalla}, {Singer}, {Sladen}, {Sooley},
  {Sornarajah}, {Streicher}, {Teuben}, {Thomas}, {Tremblay}, {Turner},
  {Terr{\'o}n}, {van Kerkwijk}, {de la Vega}, {Watkins}, {Weaver}, {Whitmore},
  {Woillez}, {Zabalza}, \& {Astropy Contributors}}]{2018AJ....156..123A}
{Astropy Collaboration}, {Price-Whelan}, A.~M., {Sip{\H{o}}cz}, B.~M., {et~al.}
  2018, \aj, 156, 123, \dodoi{10.3847/1538-3881/aabc4f}

\bibitem[{{Beck} {et~al.}(2017){Beck}, {Lin}, {Ishida}, {Gieseke}, {de Souza},
  {Costa-Duarte}, {Hattab}, \& {Krone-Martins}}]{2017MNRAS.468.4323B}
{Beck}, R., {Lin}, C.~A., {Ishida}, E.~E.~O., {et~al.} 2017, \mnras, 468, 4323,
  \dodoi{10.1093/mnras/stx687}

\bibitem[{{Bolzonella} {et~al.}(2000){Bolzonella}, {Miralles}, \&
  {Pell{\'o}}}]{2000A&A...363..476B}
{Bolzonella}, M., {Miralles}, J.~M., \& {Pell{\'o}}, R. 2000, \aap, 363, 476.
\newblock \doarXiv{astro-ph/0003380}

\bibitem[{{Brammer} {et~al.}(2008){Brammer}, {van Dokkum}, \&
  {Coppi}}]{2008ApJ...686.1503B}
{Brammer}, G.~B., {van Dokkum}, P.~G., \& {Coppi}, P. 2008, \apj, 686, 1503,
  \dodoi{10.1086/591786}

\bibitem[{{Calzetti} {et~al.}(2000){Calzetti}, {Armus}, {Bohlin}, {Kinney},
  {Koornneef}, \& {Storchi-Bergmann}}]{2000ApJ...533..682C}
{Calzetti}, D., {Armus}, L., {Bohlin}, R.~C., {et~al.} 2000, \apj, 533, 682,
  \dodoi{10.1086/308692}

\bibitem[{Crenshaw(2021)}]{Crenshaw_2021_pzflow}
Crenshaw, J.~F. 2021, {jfcrenshaw/pzflow}, v2.0.0,
  \dodoi{10.5281/zenodo.4679913}

\bibitem[{{Cunha} {et~al.}(2014){Cunha}, {Huterer}, {Lin}, {Busha}, \&
  {Wechsler}}]{2014MNRAS.444..129C}
{Cunha}, C.~E., {Huterer}, D., {Lin}, H., {Busha}, M.~T., \& {Wechsler}, R.~H.
  2014, \mnras, 444, 129, \dodoi{10.1093/mnras/stu1424}

\bibitem[{{Dalmasso} {et~al.}(2020){Dalmasso}, {Pospisil}, {Lee}, {Izbicki},
  {Freeman}, \& {Malz}}]{2020A&C....3000362D}
{Dalmasso}, N., {Pospisil}, T., {Lee}, A.~B., {et~al.} 2020, Astronomy and
  Computing, 30, 100362, \dodoi{10.1016/j.ascom.2019.100362}

\bibitem[{{DeRose} {et~al.}(2019){DeRose}, {Wechsler}, {Becker}, {Busha},
  {Rykoff}, {MacCrann}, {Erickson}, {Evrard}, {Kravtsov}, {Gruen}, {Allam},
  {Avila}, {Bridle}, {Brooks}, {Buckley-Geer}, {Carnero Rosell}, {Carrasco
  Kind}, {Carretero}, {Castander}, {Cawthon}, {Crocce}, {da Costa}, {Davis},
  {De Vicente}, {Dietrich}, {Doel}, {Drlica-Wagner}, {Fosalba}, {Frieman},
  {Garcia-Bellido}, {Gutierrez}, {Hartley}, {Hollowood}, {Hoyle}, {James},
  {Krause}, {Kuehn}, {Kuropatkin}, {Lima}, {Maia}, {Menanteau}, {Miller},
  {Miquel}, {Ogando}, {Plazas Malag{\'o}n}, {Romer}, {Sanchez}, {Schindler},
  {Serrano}, {Sevilla-Noarbe}, {Smith}, {Suchyta}, {Swanson}, {Tarle}, \&
  {Vikram}}]{2019arXiv190102401D}
{DeRose}, J., {Wechsler}, R.~H., {Becker}, M.~R., {et~al.} 2019, arXiv
  e-prints, arXiv:1901.02401.
\newblock \doarXiv{1901.02401}

\bibitem[{{Duncan} {et~al.}(2018){Duncan}, {Jarvis}, {Brown}, \&
  {R{\"o}ttgering}}]{2018MNRAS.477.5177D}
{Duncan}, K.~J., {Jarvis}, M.~J., {Brown}, M. J.~I., \& {R{\"o}ttgering}, H.
  J.~A. 2018, \mnras, 477, 5177, \dodoi{10.1093/mnras/sty940}

\bibitem[{{Etherington} {et~al.}(2017){Etherington}, {Thomas}, {Maraston},
  {Sevilla-Noarbe}, {Bechtol}, {Pforr}, {Pellegrini}, {Gschwend}, {Carnero
  Rosell}, {Maia}, {da Costa}, {Benoit-L{\'e}vy}, {Swanson}, {Hartley},
  {Abbott}, {Abdalla}, {Allam}, {Bernstein}, {Bertin}, {Brooks},
  {Buckley-Geer}, {Carrasco Kind}, {Carretero}, {Castander}, {Crocce}, {Cunha},
  {Desai}, {Doel}, {Eifler}, {Evrard}, {Fausti Neto}, {Finley}, {Flaugher},
  {Fosalba}, {Frieman}, {Gerdes}, {Gruen}, {Gruendl}, {Gutierrez}, {Honscheid},
  {James}, {Kuehn}, {Kuropatkin}, {Lahav}, {Lima}, {Martini}, {Melchior},
  {Miquel}, {Mohr}, {Nord}, {Ogando}, {Plazas}, {Romer}, {Rykoff}, {Sanchez},
  {Scarpine}, {Schubnell}, {Smith}, {Soares-Santos}, {Sobreira}, {Tarle},
  {Vikram}, {Walker}, \& {Zhang}}]{2017MNRAS.466..228E}
{Etherington}, J., {Thomas}, D., {Maraston}, C., {et~al.} 2017, \mnras, 466,
  228, \dodoi{10.1093/mnras/stw3069}

\bibitem[{{Euclid Collaboration} {et~al.}(2020){Euclid Collaboration},
  {Desprez}, {Paltani}, {Coupon}, {Almosallam}, {Alvarez-Ayllon}, {Amaro},
  {Brescia}, {Brodwin}, {Cavuoti}, {De Vicente-Albendea}, {Fotopoulou},
  {Hatfield}, {Hartley}, {Ilbert}, {Jarvis}, {Longo}, {Rau}, {Saha}, {Speagle},
  {Tramacere}, {Castellano}, {Dubath}, {Galametz}, {Kuemmel}, {Laigle},
  {Merlin}, {Mohr}, {Pilo}, {Salvato}, {Andreon}, {Auricchio}, {Baccigalupi},
  {Balaguera-Antol{\'\i}nez}, {Baldi}, {Bardelli}, {Bender}, {Biviano},
  {Bodendorf}, {Bonino}, {Bozzo}, {Branchini}, {Brinchmann}, {Burigana},
  {Cabanac}, {Camera}, {Capobianco}, {Cappi}, {Carbone}, {Carretero},
  {Carvalho}, {Casas}, {Casas}, {Castander}, {Castignani}, {Cimatti},
  {Cledassou}, {Colodro-Conde}, {Congedo}, {Conselice}, {Conversi}, {Copin},
  {Corcione}, {Courtois}, {Cuby}, {Da Silva}, {de la Torre}, {Degaudenzi}, {Di
  Ferdinando}, {Douspis}, {Duncan}, {Dupac}, {Ealet}, {Fabbian}, {Fabricius},
  {Farrens}, {Ferreira}, {Finelli}, {Fosalba}, {Fourmanoit}, {Frailis},
  {Franceschi}, {Fumana}, {Galeotta}, {Garilli}, {Gillard}, {Gillis},
  {Giocoli}, {Gozaliasl}, {Graci{\'a}-Carpio}, {Grupp}, {Guzzo}, {Hailey},
  {Haugan}, {Holmes}, {Hormuth}, {Humphrey}, {Jahnke}, {Keihanen}, {Kermiche},
  {Kilbinger}, {Kirkpatrick}, {Kitching}, {Kohley}, {Kubik}, {Kunz},
  {Kurki-Suonio}, {Ligori}, {Lilje}, {Lloro}, {Maino}, {Maiorano}, {Marggraf},
  {Markovic}, {Martinet}, {Marulli}, {Massey}, {Maturi}, {Mauri},
  {Maurogordato}, {Medinaceli}, {Mei}, {Meneghetti}, {Metcalf}, {Meylan},
  {Moresco}, {Moscardini}, {Munari}, {Niemi}, {Padilla}, {Pasian}, {Patrizii},
  {Pettorino}, {Pires}, {Polenta}, {Poncet}, {Popa}, {Potter}, {Pozzetti},
  {Raison}, {Renzi}, {Rhodes}, {Riccio}, {Rossetti}, {Saglia}, {Sapone},
  {Schneider}, {Scottez}, {Secroun}, {Serrano}, {Sirignano}, {Sirri}, {Stanco},
  {Stern}, {Sureau}, {Tallada Cresp{\'\i}}, {Tavagnacco}, {Taylor}, {Tenti},
  {Tereno}, {Toledo-Moreo}, {Torradeflot}, {Valenziano}, {Valiviita},
  {Vassallo}, {Viel}, {Wang}, {Welikala}, {Whittaker}, {Zacchei}, {Zamorani},
  {Zoubian}, \& {Zucca}}]{2020A&A...644A..31E}
{Euclid Collaboration}, {Desprez}, G., {Paltani}, S., {et~al.} 2020, \aap, 644,
  A31, \dodoi{10.1051/0004-6361/202039403}

\bibitem[{{Euclid Collaboration} {et~al.}(2021){Euclid Collaboration},
  {Ilbert}, {de la Torre}, {Martinet}, {Wright}, {Paltani}, {Laigle},
  {Davidzon}, {Jullo}, {Hildebrandt}, {Masters}, {Amara}, {Conselice},
  {Andreon}, {Auricchio}, {Azzollini}, {Baccigalupi},
  {Balaguera-Antol{\'\i}nez}, {Baldi}, {Balestra}, {Bardelli}, {Bender},
  {Biviano}, {Bodendorf}, {Bonino}, {Borgani}, {Boucaud}, {Bozzo}, {Branchini},
  {Brescia}, {Burigana}, {Cabanac}, {Camera}, {Capobianco}, {Cappi}, {Carbone},
  {Carretero}, {Carvalho}, {Casas}, {Castander}, {Castellano}, {Castignani},
  {Cavuoti}, {Cimatti}, {Cledassou}, {Colodro-Conde}, {Congedo}, {Conversi},
  {Copin}, {Corcione}, {Costille}, {Coupon}, {Courtois}, {Cropper}, {Cuby}, {Da
  Silva}, {Degaudenzi}, {Di Ferdinando}, {Dubath}, {Duncan}, {Dupac}, {Dusini},
  {Ealet}, {Fabricius}, {Farrens}, {Ferreira}, {Finelli}, {Fosalba},
  {Fotopoulou}, {Franceschi}, {Franzetti}, {Galeotta}, {Garilli}, {Gillard},
  {Gillis}, {Giocoli}, {Gozaliasl}, {Graci{\'a}-Carpio}, {Grupp}, {Guzzo},
  {Haugan}, {Holmes}, {Hormuth}, {Jahnke}, {Keihanen}, {Kermiche}, {Kiessling},
  {Kirkpatrick}, {Kunz}, {Kurki-Suonio}, {Ligori}, {Lilje}, {Lloro}, {Maino},
  {Maiorano}, {Marggraf}, {Markovic}, {Marulli}, {Massey}, {Maturi}, {Mauri},
  {Maurogordato}, {McCracken}, {Medinaceli}, {Mei}, {Metcalf}, {Moresco},
  {Morin}, {Moscardini}, {Munari}, {Nakajima}, {Neissner}, {Niemi},
  {Nightingale}, {Padilla}, {Pasian}, {Patrizii}, {Pedersen}, {Pello},
  {Pettorino}, {Pires}, {Polenta}, {Poncet}, {Popa}, {Potter}, {Pozzetti},
  {Raison}, {Renzi}, {Rhodes}, {Riccio}, {Romelli}, {Roncarelli}, {Rossetti},
  {Saglia}, {S{\'a}nchez}, {Sapone}, {Schneider}, {Schrabback}, {Scottez},
  {Secroun}, {Seidel}, {Serrano}, {Sirignano}, {Sirri}, {Stanco}, {Sureau},
  {Tallada Cresp{\'a}}, {Tenti}, {Teplitz}, {Tereno}, {Toledo-Moreo},
  {Torradeflot}, {Tramacere}, {Valentijn}, {Valenziano}, {Valiviita},
  {Vassallo}, {Wang}, {Welikala}, {Weller}, {Whittaker}, {Zacchei}, {Zamorani},
  {Zoubian}, \& {Zucca}}]{2021A&A...647A.117E}
{Euclid Collaboration}, {Ilbert}, O., {de la Torre}, S., {et~al.} 2021, \aap,
  647, A117, \dodoi{10.1051/0004-6361/202040237}

\bibitem[{{Fern{\'a}ndez-Soto} {et~al.}(2001){Fern{\'a}ndez-Soto}, {Lanzetta},
  {Chen}, {Pascarelle}, \& {Yahata}}]{2001ApJS..135...41F}
{Fern{\'a}ndez-Soto}, A., {Lanzetta}, K.~M., {Chen}, H.-W., {Pascarelle},
  S.~M., \& {Yahata}, N. 2001, \apjs, 135, 41, \dodoi{10.1086/321777}

\bibitem[{{Fontana} {et~al.}(2000){Fontana}, {D'Odorico}, {Poli}, {Giallongo},
  {Arnouts}, {Cristiani}, {Moorwood}, \& {Saracco}}]{2000AJ....120.2206F}
{Fontana}, A., {D'Odorico}, S., {Poli}, F., {et~al.} 2000, \aj, 120, 2206,
  \dodoi{10.1086/316803}

\bibitem[{{Freeman} {et~al.}(2017){Freeman}, {Izbicki}, \&
  {Lee}}]{2017MNRAS.468.4556F}
{Freeman}, P.~E., {Izbicki}, R., \& {Lee}, A.~B. 2017, \mnras, 468, 4556,
  \dodoi{10.1093/mnras/stx764}

\bibitem[{Gelman {et~al.}(2019)Gelman, Goodrich, Gabry, \&
  Vehtari}]{doi:10.1080/00031305.2018.1549100}
Gelman, A., Goodrich, B., Gabry, J., \& Vehtari, A. 2019, The American
  Statistician, 73, 307, \dodoi{10.1080/00031305.2018.1549100}

\bibitem[{{Gomes} {et~al.}(2018){Gomes}, {Jarvis}, {Almosallam}, \&
  {Roberts}}]{2018MNRAS.475..331G}
{Gomes}, Z., {Jarvis}, M.~J., {Almosallam}, I.~A., \& {Roberts}, S.~J. 2018,
  \mnras, 475, 331, \dodoi{10.1093/mnras/stx3187}

\bibitem[{Gonz{\'{a}}lez-Casanova {et~al.}(2018)Gonz{\'{a}}lez-Casanova,
  Lazarian, \& Cho}]{Gonzalez-Casanova2018}
Gonz{\'{a}}lez-Casanova, D.~F., Lazarian, A., \& Cho, J. 2018, Monthly Notices
  of the Royal Astronomical Society, 475, 3324, \dodoi{10.1093/mnras/sty006}

\bibitem[{{Hatfield} {et~al.}(2020){Hatfield}, {Almosallam}, {Jarvis}, {Adams},
  {Bowler}, {Gomes}, {Roberts}, \& {Schreiber}}]{2020MNRAS.498.5498H}
{Hatfield}, P.~W., {Almosallam}, I.~A., {Jarvis}, M.~J., {et~al.} 2020, \mnras,
  498, 5498, \dodoi{10.1093/mnras/staa2741}

\bibitem[{{Hoyle} {et~al.}(2018){Hoyle}, {Gruen}, {Bernstein}, {Rau}, {De
  Vicente}, {Hartley}, {Gaztanaga}, {DeRose}, {Troxel}, {Davis}, {Alarcon},
  {MacCrann}, {Prat}, {S{\'a}nchez}, {Sheldon}, {Wechsler}, {Asorey}, {Becker},
  {Bonnett}, {Carnero Rosell}, {Carollo}, {Carrasco Kind}, {Castander},
  {Cawthon}, {Chang}, {Childress}, {Davis}, {Drlica-Wagner}, {Gatti},
  {Glazebrook}, {Gschwend}, {Hinton}, {Hoormann}, {Kim}, {King}, {Kuehn},
  {Lewis}, {Lidman}, {Lin}, {Macaulay}, {Maia}, {Martini}, {Mudd},
  {M{\"o}ller}, {Nichol}, {Ogando}, {Rollins}, {Roodman}, {Ross}, {Rozo},
  {Rykoff}, {Samuroff}, {Sevilla-Noarbe}, {Sharp}, {Sommer}, {Tucker}, {Uddin},
  {Varga}, {Vielzeuf}, {Yuan}, {Zhang}, {Abbott}, {Abdalla}, {Allam}, {Annis},
  {Bechtol}, {Benoit-L{\'e}vy}, {Bertin}, {Brooks}, {Buckley-Geer}, {Burke},
  {Busha}, {Capozzi}, {Carretero}, {Crocce}, {D'Andrea}, {da Costa}, {DePoy},
  {Desai}, {Diehl}, {Doel}, {Eifler}, {Estrada}, {Evrard}, {Fernandez},
  {Flaugher}, {Fosalba}, {Frieman}, {Garc{\'\i}a-Bellido}, {Gerdes},
  {Giannantonio}, {Goldstein}, {Gruendl}, {Gutierrez}, {Honscheid}, {James},
  {Jarvis}, {Jeltema}, {Johnson}, {Johnson}, {Kirk}, {Krause}, {Kuhlmann},
  {Kuropatkin}, {Lahav}, {Li}, {Lima}, {March}, {Marshall}, {Melchior},
  {Menanteau}, {Miquel}, {Nord}, {O'Neill}, {Plazas}, {Romer}, {Sako},
  {Sanchez}, {Santiago}, {Scarpine}, {Schindler}, {Schubnell}, {Smith},
  {Smith}, {Soares-Santos}, {Sobreira}, {Suchyta}, {Swanson}, {Tarle},
  {Thomas}, {Tucker}, {Vikram}, {Walker}, {Weller}, {Wester}, {Wolf}, {Yanny},
  {Zuntz}, \& {DES Collaboration}}]{2018MNRAS.478..592H}
{Hoyle}, B., {Gruen}, D., {Bernstein}, G.~M., {et~al.} 2018, \mnras, 478, 592,
  \dodoi{10.1093/mnras/sty957}

\bibitem[{{Ilbert} {et~al.}(2006){Ilbert}, {Arnouts}, {McCracken},
  {Bolzonella}, {Bertin}, {Le F{\`e}vre}, {Mellier}, {Zamorani}, {Pell{\`o}},
  {Iovino}, {Tresse}, {Le Brun}, {Bottini}, {Garilli}, {Maccagni}, {Picat},
  {Scaramella}, {Scodeggio}, {Vettolani}, {Zanichelli}, {Adami}, {Bardelli},
  {Cappi}, {Charlot}, {Ciliegi}, {Contini}, {Cucciati}, {Foucaud}, {Franzetti},
  {Gavignaud}, {Guzzo}, {Marano}, {Marinoni}, {Mazure}, {Meneux}, {Merighi},
  {Paltani}, {Pollo}, {Pozzetti}, {Radovich}, {Zucca}, {Bondi}, {Bongiorno},
  {Busarello}, {de La Torre}, {Gregorini}, {Lamareille}, {Mathez}, {Merluzzi},
  {Ripepi}, {Rizzo}, \& {Vergani}}]{2006A&A...457..841I}
{Ilbert}, O., {Arnouts}, S., {McCracken}, H.~J., {et~al.} 2006, \aap, 457, 841,
  \dodoi{10.1051/0004-6361:20065138}

\bibitem[{{Jimenez Rezende} \& {Mohamed}(2015)}]{2015arXiv150505770J}
{Jimenez Rezende}, D., \& {Mohamed}, S. 2015, arXiv e-prints, arXiv:1505.05770.
\newblock \doarXiv{1505.05770}

\bibitem[{{Laureijs} {et~al.}(2011){Laureijs}, {Amiaux}, {Arduini},
  {Augu{\`e}res}, {Brinchmann}, {Cole}, {Cropper}, {Dabin}, {Duvet}, {Ealet},
  {Garilli}, {Gondoin}, {Guzzo}, {Hoar}, {Hoekstra}, {Holmes}, {Kitching},
  {Maciaszek}, {Mellier}, {Pasian}, {Percival}, {Rhodes}, {Saavedra Criado},
  {Sauvage}, {Scaramella}, {Valenziano}, {Warren}, {Bender}, {Castander},
  {Cimatti}, {Le F{\`e}vre}, {Kurki-Suonio}, {Levi}, {Lilje}, {Meylan},
  {Nichol}, {Pedersen}, {Popa}, {Rebolo Lopez}, {Rix}, {Rottgering},
  {Zeilinger}, {Grupp}, {Hudelot}, {Massey}, {Meneghetti}, {Miller}, {Paltani},
  {Paulin-Henriksson}, {Pires}, {Saxton}, {Schrabback}, {Seidel}, {Walsh},
  {Aghanim}, {Amendola}, {Bartlett}, {Baccigalupi}, {Beaulieu}, {Benabed},
  {Cuby}, {Elbaz}, {Fosalba}, {Gavazzi}, {Helmi}, {Hook}, {Irwin}, {Kneib},
  {Kunz}, {Mannucci}, {Moscardini}, {Tao}, {Teyssier}, {Weller}, {Zamorani},
  {Zapatero Osorio}, {Boulade}, {Foumond}, {Di Giorgio}, {Guttridge}, {James},
  {Kemp}, {Martignac}, {Spencer}, {Walton}, {Bl{\"u}mchen}, {Bonoli},
  {Bortoletto}, {Cerna}, {Corcione}, {Fabron}, {Jahnke}, {Ligori}, {Madrid},
  {Martin}, {Morgante}, {Pamplona}, {Prieto}, {Riva}, {Toledo}, {Trifoglio},
  {Zerbi}, {Abdalla}, {Douspis}, {Grenet}, {Borgani}, {Bouwens}, {Courbin},
  {Delouis}, {Dubath}, {Fontana}, {Frailis}, {Grazian}, {Koppenh{\"o}fer},
  {Mansutti}, {Melchior}, {Mignoli}, {Mohr}, {Neissner}, {Noddle}, {Poncet},
  {Scodeggio}, {Serrano}, {Shane}, {Starck}, {Surace}, {Taylor},
  {Verdoes-Kleijn}, {Vuerli}, {Williams}, {Zacchei}, {Altieri}, {Escudero
  Sanz}, {Kohley}, {Oosterbroek}, {Astier}, {Bacon}, {Bardelli}, {Baugh},
  {Bellagamba}, {Benoist}, {Bianchi}, {Biviano}, {Branchini}, {Carbone},
  {Cardone}, {Clements}, {Colombi}, {Conselice}, {Cresci}, {Deacon}, {Dunlop},
  {Fedeli}, {Fontanot}, {Franzetti}, {Giocoli}, {Garcia-Bellido}, {Gow},
  {Heavens}, {Hewett}, {Heymans}, {Holland}, {Huang}, {Ilbert}, {Joachimi},
  {Jennins}, {Kerins}, {Kiessling}, {Kirk}, {Kotak}, {Krause}, {Lahav}, {van
  Leeuwen}, {Lesgourgues}, {Lombardi}, {Magliocchetti}, {Maguire}, {Majerotto},
  {Maoli}, {Marulli}, {Maurogordato}, {McCracken}, {McLure}, {Melchiorri},
  {Merson}, {Moresco}, {Nonino}, {Norberg}, {Peacock}, {Pello}, {Penny},
  {Pettorino}, {Di Porto}, {Pozzetti}, {Quercellini}, {Radovich}, {Rassat},
  {Roche}, {Ronayette}, {Rossetti}, {Sartoris}, {Schneider}, {Semboloni},
  {Serjeant}, {Simpson}, {Skordis}, {Smadja}, {Smartt}, {Spano}, {Spiro},
  {Sullivan}, {Tilquin}, {Trotta}, {Verde}, {Wang}, {Williger}, {Zhao},
  {Zoubian}, \& {Zucca}}]{2011arXiv1110.3193L}
{Laureijs}, R., {Amiaux}, J., {Arduini}, S., {et~al.} 2011, arXiv e-prints,
  arXiv:1110.3193.
\newblock \doarXiv{1110.3193}

\bibitem[{Lewis-beck \& Skalaban(1990)}]{Lewis-beck1990}
Lewis-beck, M.~S., \& Skalaban, A. 1990, Political Analysis, 2, 153,
  \dodoi{10.1093/PAN/2.1.153}

\bibitem[{{LSST Science Collaboration} {et~al.}(2009){LSST Science
  Collaboration}, {Abell}, {Allison}, {Anderson}, {Andrew}, {Angel}, {Armus},
  {Arnett}, {Asztalos}, {Axelrod}, {Bailey}, {Ballantyne}, {Bankert},
  {Barkhouse}, {Barr}, {Barrientos}, {Barth}, {Bartlett}, {Becker}, {Becla},
  {Beers}, {Bernstein}, {Biswas}, {Blanton}, {Bloom}, {Bochanski}, {Boeshaar},
  {Borne}, {Bradac}, {Brandt}, {Bridge}, {Brown}, {Brunner}, {Bullock},
  {Burgasser}, {Burge}, {Burke}, {Cargile}, {Chandrasekharan}, {Chartas},
  {Chesley}, {Chu}, {Cinabro}, {Claire}, {Claver}, {Clowe}, {Connolly}, {Cook},
  {Cooke}, {Cooray}, {Covey}, {Culliton}, {de Jong}, {de Vries}, {Debattista},
  {Delgado}, {Dell'Antonio}, {Dhital}, {Di Stefano}, {Dickinson}, {Dilday},
  {Djorgovski}, {Dobler}, {Donalek}, {Dubois-Felsmann}, {Durech},
  {Eliasdottir}, {Eracleous}, {Eyer}, {Falco}, {Fan}, {Fassnacht}, {Ferguson},
  {Fernandez}, {Fields}, {Finkbeiner}, {Figueroa}, {Fox}, {Francke}, {Frank},
  {Frieman}, {Fromenteau}, {Furqan}, {Galaz}, {Gal-Yam}, {Garnavich},
  {Gawiser}, {Geary}, {Gee}, {Gibson}, {Gilmore}, {Grace}, {Green}, {Gressler},
  {Grillmair}, {Habib}, {Haggerty}, {Hamuy}, {Harris}, {Hawley}, {Heavens},
  {Hebb}, {Henry}, {Hileman}, {Hilton}, {Hoadley}, {Holberg}, {Holman},
  {Howell}, {Infante}, {Ivezic}, {Jacoby}, {Jain}, {R}, {Jedicke}, {Jee},
  {Garrett Jernigan}, {Jha}, {Johnston}, {Jones}, {Juric}, {Kaasalainen},
  {Styliani}, {Kafka}, {Kahn}, {Kaib}, {Kalirai}, {Kantor}, {Kasliwal},
  {Keeton}, {Kessler}, {Knezevic}, {Kowalski}, {Krabbendam}, {Krughoff},
  {Kulkarni}, {Kuhlman}, {Lacy}, {Lepine}, {Liang}, {Lien}, {Lira}, {Long},
  {Lorenz}, {Lotz}, {Lupton}, {Lutz}, {Macri}, {Mahabal}, {Mandelbaum},
  {Marshall}, {May}, {McGehee}, {Meadows}, {Meert}, {Milani}, {Miller},
  {Miller}, {Mills}, {Minniti}, {Monet}, {Mukadam}, {Nakar}, {Neill}, {Newman},
  {Nikolaev}, {Nordby}, {O'Connor}, {Oguri}, {Oliver}, {Olivier}, {Olsen},
  {Olsen}, {Olszewski}, {Oluseyi}, {Padilla}, {Parker}, {Pepper}, {Peterson},
  {Petry}, {Pinto}, {Pizagno}, {Popescu}, {Prsa}, {Radcka}, {Raddick},
  {Rasmussen}, {Rau}, {Rho}, {Rhoads}, {Richards}, {Ridgway}, {Robertson},
  {Roskar}, {Saha}, {Sarajedini}, {Scannapieco}, {Schalk}, {Schindler},
  {Schmidt}, {Schmidt}, {Schneider}, {Schumacher}, {Scranton}, {Sebag},
  {Seppala}, {Shemmer}, {Simon}, {Sivertz}, {Smith}, {Allyn Smith}, {Smith},
  {Spitz}, {Stanford}, {Stassun}, {Strader}, {Strauss}, {Stubbs}, {Sweeney},
  {Szalay}, {Szkody}, {Takada}, {Thorman}, {Trilling}, {Trimble}, {Tyson}, {Van
  Berg}, {Vanden Berk}, {VanderPlas}, {Verde}, {Vrsnak}, {Walkowicz},
  {Wandelt}, {Wang}, {Wang}, {Warner}, {Wechsler}, {West}, {Wiecha},
  {Williams}, {Willman}, {Wittman}, {Wolff}, {Wood-Vasey}, {Wozniak}, {Young},
  {Zentner}, \& {Zhan}}]{2009arXiv0912.0201L}
{LSST Science Collaboration}, {Abell}, P.~A., {Allison}, J., {et~al.} 2009,
  arXiv e-prints, arXiv:0912.0201.
\newblock \doarXiv{0912.0201}

\bibitem[{Malz(2021)}]{Malz2021}
Malz, A.~I. 2021, Physical Review D, 103, 083502,
  \dodoi{10.1103/PhysRevD.103.083502}

\bibitem[{{Malz} {et~al.}(2018){Malz}, {Marshall}, {DeRose}, {Graham},
  {Schmidt}, {Wechsler}, \& {(LSST Dark Energy Science
  Collaboration}}]{2018AJ....156...35M}
{Malz}, A.~I., {Marshall}, P.~J., {DeRose}, J., {et~al.} 2018, \aj, 156, 35,
  \dodoi{10.3847/1538-3881/aac6b5}

\bibitem[{{Mitra} \& {Linder}(2021)}]{2021PhRvD.103b3524M}
{Mitra}, A., \& {Linder}, E.~V. 2021, \prd, 103, 023524,
  \dodoi{10.1103/PhysRevD.103.023524}

\bibitem[{{Miyaji} {et~al.}(2015){Miyaji}, {Hasinger}, {Salvato}, {Brusa},
  {Cappelluti}, {Civano}, {Puccetti}, {Elvis}, {Brunner}, {Fotopoulou}, {Ueda},
  {Griffiths}, {Koekemoer}, {Akiyama}, {Comastri}, {Gilli}, {Lanzuisi},
  {Merloni}, \& {Vignali}}]{2015ApJ...804..104M}
{Miyaji}, T., {Hasinger}, G., {Salvato}, M., {et~al.} 2015, \apj, 804, 104,
  \dodoi{10.1088/0004-637X/804/2/104}

\bibitem[{{Polsterer} {et~al.}(2016){Polsterer}, {D'Isanto}, \&
  {Gieseke}}]{2016arXiv160808016P}
{Polsterer}, K.~L., {D'Isanto}, A., \& {Gieseke}, F. 2016, arXiv e-prints,
  arXiv:1608.08016.
\newblock \doarXiv{1608.08016}

\bibitem[{{Salvato} {et~al.}(2019){Salvato}, {Ilbert}, \&
  {Hoyle}}]{2019NatAs...3..212S}
{Salvato}, M., {Ilbert}, O., \& {Hoyle}, B. 2019, Nature Astronomy, 3, 212,
  \dodoi{10.1038/s41550-018-0478-0}

\bibitem[{{Schmidt} {et~al.}(2020){Schmidt}, {Malz}, {Soo}, {Almosallam},
  {Brescia}, {Cavuoti}, {Cohen-Tanugi}, {Connolly}, {DeRose}, {Freeman},
  {Graham}, {Iyer}, {Jarvis}, {Kalmbach}, {Kovacs}, {Lee}, {Longo}, {Morrison},
  {Newman}, {Nourbakhsh}, {Nuss}, {Pospisil}, {Tranin}, {Wechsler}, {Zhou},
  {Izbicki}, \& {LSST Dark Energy Science Collaboration}}]{2020MNRAS.499.1587S}
{Schmidt}, S.~J., {Malz}, A.~I., {Soo}, J.~Y.~H., {et~al.} 2020, \mnras, 499,
  1587, \dodoi{10.1093/mnras/staa2799}

\bibitem[{Shoji {et~al.}(2020)Shoji, Takata, \& Mizumoto}]{Shoji2020}
Shoji, I., Takata, T., \& Mizumoto, Y. 2020, Monthly Notices of the Royal
  Astronomical Society, 495, 338, \dodoi{10.1093/MNRAS/STAA1159}

\bibitem[{{Simm} {et~al.}(2015){Simm}, {Saglia}, {Salvato}, {Bender},
  {Burgett}, {Chambers}, {Draper}, {Flewelling}, {Kaiser}, {Kudritzki},
  {Magnier}, {Metcalfe}, {Tonry}, {Wainscoat}, \&
  {Waters}}]{2015A&A...584A.106S}
{Simm}, T., {Saglia}, R., {Salvato}, M., {et~al.} 2015, \aap, 584, A106,
  \dodoi{10.1051/0004-6361/201526859}

\bibitem[{{Tanaka} {et~al.}(2018){Tanaka}, {Coupon}, {Hsieh}, {Mineo},
  {Nishizawa}, {Speagle}, {Furusawa}, {Miyazaki}, \&
  {Murayama}}]{2018PASJ...70S...9T}
{Tanaka}, M., {Coupon}, J., {Hsieh}, B.-C., {et~al.} 2018, \pasj, 70, S9,
  \dodoi{10.1093/pasj/psx077}

\bibitem[{{The Dark Energy Survey Collaboration}(2005)}]{2005astro.ph.10346T}
{The Dark Energy Survey Collaboration}. 2005, arXiv e-prints, astro.
\newblock \doarXiv{astro-ph/0510346}

\bibitem[{Wright(1921)}]{Wright1921}
Wright, S. 1921, Journal of Agricultural Research, 20, 557

\bibitem[{Zhao {et~al.}(2021)Zhao, Dalmasso, Izbicki, \&
  Lee}]{zhao2021diagnostics}
Zhao, D., Dalmasso, N., Izbicki, R., \& Lee, A.~B. 2021, in Proceedings of
  Machine Learning Research, Vol. 125, Proceedings of the 37th Conference on
  Uncertainty in Artificial Intelligence (UAI) (PMLR)

\bibitem[{{Zuntz} {et~al.}(2021){Zuntz}, {Lanusse}, {Malz}, {Wright}, {Slosar},
  {Abolfathi}, {Alonso}, {Bault}, {Bom}, {Brescia}, {Broussard}, {Campagne},
  {Cavuoti}, {Cypriano}, {Fraga}, {Gawiser}, {Gonzalez}, {Green}, {Hatfield},
  {Iyer}, {Kirkby}, {Nicola}, {Nourbakhsh}, {Park}, {Teixeira}, {Heitmann},
  {Kovacs}, {Mao}, \& {LSST Dark Energy Science
  Collaboration}}]{2021OJAp....4E..13Z}
{Zuntz}, J., {Lanusse}, F., {Malz}, A.~I., {et~al.} 2021, The Open Journal of
  Astrophysics, 4, 13, \dodoi{10.21105/astro.2108.13418}

\end{thebibliography}
\bibliographystyle{aasjournal}

\end{document}